\documentclass[twocolumn,twocolappendix]{aastex631}
\usepackage{graphicx}
\usepackage{float}
\usepackage{newtxtext,newtxmath}
\usepackage{pifont}

\newcommand{\nut}{{\sc nut}}

\newcommand{\MBinj}{\textbf{MBinj}}

\newcommand{\K}{\mathrm{K}}

\newcommand{\s}{\mathrm{s}}
\newcommand{\g}{\mathrm{g}}
\newcommand{\cm}{\mathrm{cm}}
\newcommand{\cc}{\mathrm{cm}^{-3}}

\newcommand{\kpc}{\mathrm{kpc}}

\newcommand{\Msun}{\mathrm{M}_\odot}
\newcommand{\ramses}{{\sc ramses}}
\shorttitle{the fallibility of equipartition magnetic field strengths}
\shortauthors{dacunha et al.}

\begin{document}

\title{The fallibility of equipartition magnetic field strengths from synchrotron emission using synthetically observed galaxies}

\author[0000-0002-4746-2128]{Tara Dacunha}
\affiliation{Department of Physics, Stanford University, Stanford, California 94305, USA}
\affiliation{Kavli Institute for Particle Astrophysics \& Cosmology (KIPAC), Stanford University, Stanford, CA 94305, USA}

\author[0000-0002-4059-9850]{Sergio Martin-Alvarez}
\affiliation{Kavli Institute for Particle Astrophysics \& Cosmology (KIPAC), Stanford University, Stanford, CA 94305, USA}

\author[0000-0002-7633-3376]{Susan E. Clark}
\affiliation{Department of Physics, Stanford University, Stanford, California 94305, USA}
\affiliation{Kavli Institute for Particle Astrophysics \& Cosmology (KIPAC), Stanford University, Stanford, CA 94305, USA}

\author[0000-0001-5357-6538]{Enrique Lopez-Rodriguez}
\affiliation{Kavli Institute for Particle Astrophysics \& Cosmology (KIPAC), Stanford University, Stanford, CA 94305, USA}
\affiliation{Department of Physics \& Astronomy, University of South Carolina, Columbia, SC 29208, USA}

\email{tdacunha@stanford.edu}









\begin{abstract}
Understanding the role that magnetic fields play on the stage of galaxy formation requires accurate methods for inferring the properties of extragalactic magnetic fields. Radio synchrotron emission has been the most promising avenue to infer magnetic field strengths across galaxies, with the application of a central assumption: that galactic cosmic rays are in energy equipartition with the magnetic field. In this work, we leverage flexible synthetic observations of a high-resolution magnetohydrodynamic simulation of a Milky Way-like galaxy to review whether true equipartition is capable of reproducing radio observations of galaxies, and investigate its impact on the inference of magnetic field strengths when varying the properties and density distribution of the cosmic rays. We find that imposing equipartition (regardless of scale length) results in cosmic ray electron densities that are unable to generate either the amplitude or the shape of the radio intensity profiles typically observed in spiral galaxies. Instead, observationally motivated smooth distributions of cosmic ray electrons across the galaxy provide a remarkable match to observations. We further demonstrate that assuming equipartition with those mock observations can lead to significant overestimation of the magnetic field strength. This misestimation varies with cosmic ray electron densities, cosmic ray spectrum power-law index, and galactic environment, aggravated in inter-arm regions and attenuated in star-forming regions. Our results promote caution when assuming equipartition in observations, and suggest that additional theoretical and numerical work is required to leverage the upcoming generation of radio observations poised to revolutionize our understanding of astrophysical magnetic fields.
\end{abstract}

\keywords{Astrophysical magnetism (102); Extragalactic magnetic fields (507); Radio continuum emission (1340); Spiral galaxies (1560); Disk galaxies (391); Astronomical simulations (1857); Magnetohydrodynamical simulations (1966); Cosmic rays (329)}



\section{Introduction}

Magnetic fields are as fundamental across astrophysics as it is complex to measure them. In the context of galaxies, they are pervasive across most galaxy types \citep{Chyzy2011, Beck2015, Basu2017, Lopez-Rodriguez2022b} and across time \citep{Bernet2008, Mao2017, Geach2023, Chen2024}. They are essential to the evolution of the interstellar medium (ISM) and crucial for several of the processes it harbors. These magnetic fields play a key role in co-regulating star formation \citep[e.g.,][]{McKeeOstriker2007, Federrath2012, Robinson2023}, providing gas pressure support against gravitational contraction \citep{Boulares1990,Kortgen2019, Chang-Goo2023}, modifying the gas distribution across phases in the ISM \citep{Iffrig2017, Hennebelle2019}, governing the propagation of cosmic rays \citep{Kulsrud1969, Wentzel1974, Zweibel2013, Shukurov2017, Iryna2024}, suppressing galactic outflows \citep{Bendre2015, Shukurov2018}, and even affecting the hierarchical growth of galaxies \citep[e.g.,][]{Pillepich2018a,Martin-Alvarez2020,Whittingham2021, Sanati2024}. Even so, much about the interplay between magnetic fields, the structure of the ISM, and the evolution of galaxies remains to be disentangled. 

In order to probe these questions and build our understanding of the properties and impact of the magnetic field, detailed and accurate observations of its strength and spatial distribution are required. 
While acquiring such information about galaxies is nontrivial, there are a number of observational avenues that provide complementary clues to the underlying magnetic field configuration. These observational signatures include total synchrotron intensity, polarized synchrotron intensity, Faraday depolarization, Faraday rotation, the longitudinal Zeeman effect, and optical, infrared, and submillimeter polarization \citep{BeckReview2015}.

Particularly relevant for our work are observations of synchrotron emission in the radio band. It is particularly useful in face-on observations of spiral galaxies as the intensity of this radiation probes the strength of the component of the magnetic field perpendicular to the line-of-sight \citep{BeckReview2015}. The fact that virtually all galaxies with some degree of ongoing star formation emit this synchrotron radiation (see \citet{Condon1992} for a review of radio emission in galaxies) reflects the pervasive presence of magnetic fields in all galaxies. Furthermore, the polarization of this synchrotron emission provides information about the structure of magnetic fields across the ISM of galaxies, probing the large-scale spiral magnetic field attributed to the disk's differential rotation in face-on observations \citep{Beck2007, Mao2008, Beck2015, Berkhuijsen2016, Lopez-Rodriguez2020NGC1068, Borlaff2023ExtragalacticGalaxies} as well as the magnetic fields across the galaxy-halo interface in edge-on observations \citep{Krause2018_CHANG-ES, Mulcahy2018, Mora-Partiarroyo2019CHANG-ES4631}.

Both from a theoretical \citep{tomography2024} and observational \citep{Borlaff2023ExtragalacticGalaxies} perspective, accurate, panchromatic studies are fundamental to understanding the complex, multi-scale configuration of magnetic fields in galaxies. The complementarity between different emission types is driven by the properties and distribution across the ISM of their generating elements, posing them as windows to different scales and ISM phases \citep{tomography2024}. In the context of synchrotron, this emission is radiated by cosmic ray (CR) electrons gyrating in galactic magnetic fields. Therefore its intensity is modulated both by the magnetic field strength perpendicular to the line-of-sight, $B_{\rm tot, \perp}$, and the CR electron number density and energy spectrum. 
To estimate the magnetic field strength perpendicular to the line-of-sight, it is therefore necessary to either obtain supplementary information characterizing the CR electrons, or to make assumptions about the CR distribution. This can be done, e.g., by relating the CR properties to those of the {\it inferred} magnetic field. Extragalactic CRs are notoriously difficult if not impossible to measure directly and localize to their sources due to energy losses and the deflection of the remaining high energy CRs by interceding magnetic fields \citep{Grenier2015}. This makes external information about their density distribution exceedingly rare \citep{Seta&Beck2019}. Consequently, a reasonable assumption to characterize their properties is that of energy equipartition between the energy carried by these relativistic particles and that contained in the magnetic field. Such an assumption has proven a powerful and even necessary avenue for estimating the magnetic field strengths in many radio sources \citep{Pacholczyk1970}. Using this assumption in radio polarimetric observations with spatial resolutions of $\sim$100s pcs, the total magnetic field strength in spiral galaxies is estimated to be $\sim17\pm14~\mu$G \citep{Fletcher2010,Beck2019}. Works including \citet{Beck2000}, \citet{Fletcher2011}, \citet{Soida2011}, \citet{Lacki2013TheLosses}, \citet{BasuRoy2013}, and \citet{Krause2018_CHANG-ES} have reported equipartition estimates of the magnetic field strengths of spiral and other galaxies, most using the revised equipartition formulae derived in \citet{Beck2005RevisedObservations}. In many cases, equipartition field estimates can be compared to other tracers of the magnetic field strengths such as Faraday rotation measure, Zeeman splitting, and Voyager in-situ measurements \citep[e.g.,][]{Haverkorn2015, Beck2019, Voyager_future}. These tracers allow us to probe different components of the magnetic field, phases of the ISM, and physical scales.

The equipartition assumption provides similar results to the minimum-energy assumption which was proposed by \citet{Burbidge1956} in the context of optical synchrotron emission from the M87 jet. Both assumptions rely on the strong coupling of the magnetic field and the CRs \citep{Beck2005RevisedObservations}. Magnetic fields in spiral galaxies are expected to be within an order of magnitude of the equipartition estimate since overly weak magnetic fields would allow the escape of most CR particles, resulting in a lack of synchrotron emission. Conversely, overly strong magnetic fields would confine CRs close to their production sources due to short propagation lengths, defying observations of the thick and extended synchrotron disks of galaxies \citep{Duric, Krause2018_CHANG-ES}. 

However, as useful as it is in facilitating the measurement of magnetic field strengths, there has been growing scrutiny of the validity of this commonplace equipartition assumption.  \citet{Stepanov2014} determined using observations and modelling of synchrotron emission from the Milky Way and the spiral galaxy M33 that equipartition does not seem to hold on scales smaller than 1 kpc. \citet{Yoast-Hull2016} used gamma-ray and radio spectra of the central molecular zones of nearby starburst galaxies to conclude that equipartition is an unreliable assumption in regions of such intense star formation. 
\citet{Seta2018} and \citet{Seta&Beck2019} have raised concerns about the theoretical justification for the equipartition assumption and claimed that below the driving scale of supernova-driven turbulence, around 100 pc, CR and magnetic field energy densities are not correlated, making equipartition invalid below those scales. 

With the entangled unknowns of CRs and magnetic fields in synchrotron observations, magnetohydrodynamic (MHD) simulations of galaxy formation provide an invaluable avenue for probing the implications of the equipartition assumption. High-resolution MHD simulations have expanded in the last decade and are able to simulate galaxies with realistic magnetic properties and magnetic amplification processes (e.g., \citet{Dubois2010, Pakmor2017, Hopkins2020, Martin-Alvarez2021, Martin-Alvarez2023}). 
\citet{Werhan2021_I} used MHD simulations of galaxies including on-the-fly $\sim$GeV CR proton energy density modeling combined with post-processing to determine the steady-state spectra of the CR protons and electrons accounting for various loss mechanisms. They further used this method to probe underlying physics involved in far-infrared and radio emission over a range of star formation rates \citep{Werhahn2021_III}. 
\citet{Ponnada2024} used MHD galaxy formation simulations of three Milky Way-like galaxies with a model for spectrally-resolved CRs \citep{Hopkins2022} to probe the equipartition assumption and its relation to the volumes and phases responsible for synchrotron emission. They found nonnegligible discrepancies between their intrinsic magnetic field strengths and those measured by combining the synchrotron emission with the standard equipartition assumptions, leading to magnetic field strength underestimation.

In this work, we leverage flexible synthetic observations of a constrained transport, high-resolution MHD zoom-in simulation of a Milky Way-like galaxy to gauge the impact of the energy equipartition assumption on the recovery of magnetic field strengths from observed synchrotron intensities over a range of CR properties.
We focus on equipartition between the CR protons (which dominate the energy budget) and the magnetic field (``proton equipartition"). However, we also briefly consider the effects of a less common equipartition assumption, directly between the magnetic energy and the radiating electrons (``electron equipartition").
To probe the impact of an equipartition assumption, we test two overall models for the spatial density distribution of the CRs, a prescribed smooth exponential distribution and a distribution required by imposing equipartition with the simulated magnetic field of the galaxy. We examine the resulting synthetically observed intensities and the deviations between the subsequently inferred and intrinsic magnetic field strength as a function of the CR energy spectrum power-law index, CR electron density, and spatial region in the galaxy.  

The structure of this work is as follows: the high-resolution MHD simulations and numerical methods are presented in \autoref{sec:sim}. The methods and CR distribution models used to generate our synthetic observations are discussed in \autoref{sec:synth_obs}. The equipartition magnetic field equations are derived and discussed in \autoref{sec:EqB}. Our results assuming CR proton equipartition are presented and analyzed in \autoref{sec:results}. A discussion on the treatment of the CR electron spectrum and on spatial variations in the spectrum power-law index is included in \autoref{sec:discussion}. We summarize our work and conclude in \autoref{sec:conclusions}.

\section{Numerical Methods}
\label{sec:sim}

The numerical simulations analyzed in this work have been previously presented and studied in \citet{Martin-Alvarez2020}, \citet{Martin-Alvarez2021}, and \citet{tomography2024}. Our MHD simulations have been generated employing our own modified version of the public code \ramses~\citep{Teyssier2002}. \ramses~ employs an Eulerian treatment of the gas along with collisionless particles for the stellar and dark matter components, all coupled by a gravity solver. \ramses~makes use of a discretized adaptive mesh refinement (AMR) octree grid to evolve the gas fluid hydrodynamics and, crucially, implements a constrained transport (CT) treatment of the magnetic field that ensures the magnetic field solenoidality by construction \citep{Teyssier2006, Fromang2006}. For further details regarding the importance of CT in galaxies, see \citet{tomography2024}.

\subsection{Initial Conditions}
\label{ss:ICs}

Our high-resolution cosmological simulations study the "\nut~galaxy", a Milky Way-like system forming in the centre of an approximately spherical zoom region. These initial conditions (ICs) were originally presented by \citet{Powell2011}. The galaxy formation and evolution of  the \nut~galaxy has been studied in detail over the last decade \citep{Geen2012, Tillson2015, Kimm2017,Martin-Alvarez2018, Rodriguez-Montero2023}.

The \nut~galaxy is simulated in a cubic box with 12.5 comoving Mpc (cMpc) per side, with a spherical zoom region approximately $4.5$~cMpc across. The \nut~galaxy resides in a dark matter halo with virial mass $M_\text{vir} (z = 0) \simeq 5 \cdot 10^{11} \Msun$. The mass resolution for the dark matter ($m_\text{DM}$) and stellar components ($m_{*}$) are $m_\text{DM} \simeq 5 \cdot 10^4 \Msun$ and $m_{*} \simeq 5 \cdot 10^3 \Msun$, respectively. We allow AMR spatial refinement down to a minimum physical cell size $\sim10$~pc (equivalent to $\sim5$~pc for a particle-like treatment). At this resolution, \nut~showcases turbulent dynamo amplification \citep{Martin-Alvarez2018, Martin-Alvarez2022}. All our simulations are generated assuming a WMAP5 cosmology \citep{Dunkley2009}.

\subsection{Galaxy Formation Physics}
\label{ss:Subgrid}

\begin{figure}[t]
\includegraphics[width=\columnwidth]{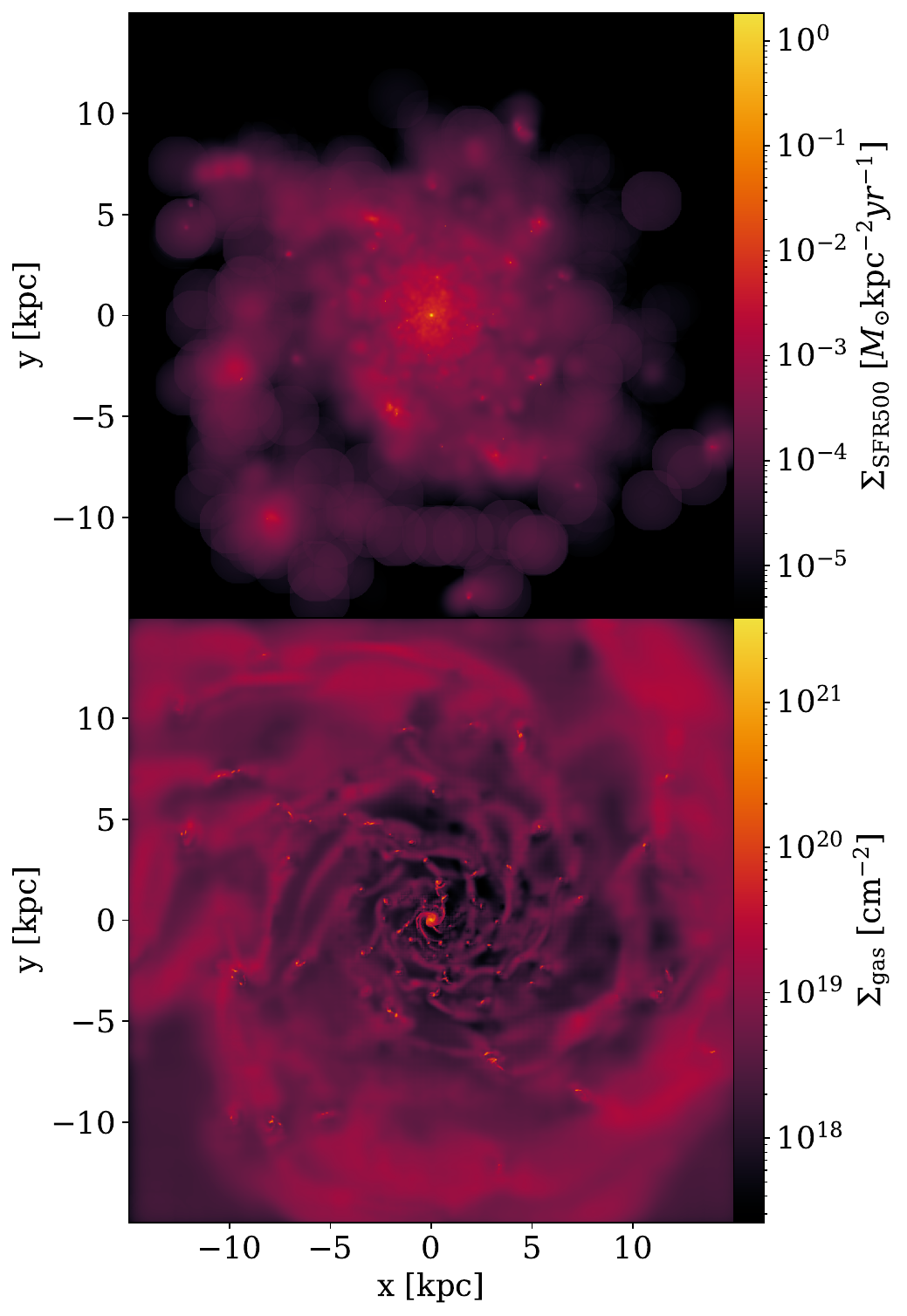}
    \caption{Face-on view of the star formation rate surface density averaged over the last 500 Myr (top) and the total gas surface number density (bottom) of the \nut~galaxy as seen at $z = 0.95$.}
    \label{fig:intrinsic_maps}
\end{figure}
To model the formation of a realistic galaxy, we include additional physical prescriptions briefly described in this section.
We model gas cooling accounting for the metal content and thermodynamical properties of each gas cell. For temperatures above $10^4$~K we interpolate {\sc cloudy} cooling tables \citep{Ferland1998}. Crucially for the formation of a realistic ISM, we also model cooling below $10^4$~K following \citet{Rosen1995}. This specific modeling of gas cooling at low temperatures is necessary for a self-consistent treatment of the cold gas distribution within the ISM. We also include a uniform UV background, turned on at $z = 10$, in order to model the impact of reionization \citep{Haardt1996}. 

We model star formation through a magneto-thermo-turbulent (MTT) star formation prescription \citet{Kimm2017, Trebitsch2017, Martin-Alvarez2020}. It allows star formation in cells at the highest level of refinement in which the gravitational pull is larger than the MTT pressure support, and employing a local star formation efficiency \citep{Federrath2012}. We refer to Appendix B of \citet{Martin-Alvarez2020} for details on the implementation of MTT star formation.

All stellar particles in our simulation are allowed to generate supernova (SN) events following the mechanical stellar feedback prescription presented by \citet{Kimm2014, Kimm2015}. Each stellar particle injects mass, momentum and energy back to its hosting cell and its immediate neighbours. For further details on the models governing the number and behavior of these SN events, we refer to \citet{tomography2024}.

In \autoref{fig:intrinsic_maps}, we offer a face-on view of two intrinsic properties of the simulated galaxy with ongoing star formation in dense gas clumps: the star formation rate surface density and the total gas surface number density. In this work, the galaxy is studied at redshift $z = 0.95$. The system is a stable disk galaxy matching the properties of late-type spiral galaxies and has star formation predominantly taking place within a radius of 8 kpc.

\subsection{Magnetic Field Models}
\label{ss:Suite}

We mainly focus on a model of magnetization where the galaxy is magnetized by ab-initio magnetic fields described by $B_0$, the comoving strength seeded uniformly at the start of the simulation ($z = 499$). We focus on the $B_0 \sim 10^{-11}$ G model (well-below the current Planck constraints of $10^{-9}$ \citep{PlanckCollaboration2015}). See \citet{tomography2024} for further details on this magnetized run of the galaxy as well as other $B_0$ models studied. We also have also reviewed the other models in \citet{tomography2024}, and analyzed in detail a model where magnetic fields are sourced by magnetized SN feedback (i.e., injecting 1\% of each SN in the form of magnetic energy; \MBinj). While the results are not included here, they confirm the main findings of this work.

\section{Synthetic radio observations and Cosmic Ray Electron Models}
\label{sec:synth_obs}

Our synthetic synchrotron observations for radio frequencies of the simulated galaxy are generated with our modified version of the {\sc polaris} code, a radiative transfer OpenMP parallelized code \citep{Reissl2016,Reissl2019}. Our modifications include some minor corrections to the synchrotron module made available on 2023/07/31 at its public repository (\href{https://github.com/polaris-MCRT/POLARIS}{https://github.com/polaris-MCRT/POLARIS}\footnote{The public website for {\sc polaris} is: \href{https://portia.astrophysik.uni-kiel.de/polaris/}{https://portia.astrophysik.uni-kiel.de/polaris/}}). To focus the scope of our investigation, we limit our study to emission at wavelength $\lambda = 6.2$ cm, and leave multi-frequency studies for future work. This wavelength corresponds to one of the main radio telescopes observing bands, targeted both by current observatories such as the Very Large Array (VLA), the Effelsberg telescope or MeerKAT; and by the upcoming Mid array from Square Kilometre Array
 (SKA). Our main comparison is with the observations of the spiral galaxy IC~342 presented by \citet{Beck2015}, which combine data from VLA and Effelsberg in this wavelength. To interface our simulations with {\sc polaris}, we process the simulation with our {\sc ramses2polaris} code. Note that we do not modify or post-process the magnetic field from the simulations  \citep[e.g.,][]{Reissl2019}. Instead, we provide {\sc polaris} with the unmodified magnetic field from the simulations, employing full-resolution information from the entire AMR grid. As a result, the capture of small-scale structure of the gas and magnetic
turbulence is only limited by resolution \citep{Kortgen2017, Martin-Alvarez2022}. We provide further detail about this interfacing in \citet{tomography2024}.

Besides the full 3D configuration of the magnetic field, generating synthetic synchrotron observations with {\sc polaris} requires information on the distribution of CR electrons, which is not directly modeled in the simulation. We model the CR electrons energy spectrum as a decreasing power-law with index $p$, which we vary from 2.2 to 3.0. The lower values of this range correspond to flatter power-law indices, expected in starburst galaxies and in star forming regions with recent CR acceleration. The higher values correspond to old spectra of electrons having experienced energy losses \citep{Lacki2013TheLosses, Beck2015}. Similarly to \citet{Reissl2019}, we consider a kinetic energy range for the spectrum bounded by Lorentz factors:
\begin{equation}
    \label{eq:gamma}
    \gamma \in [5, 301].
\end{equation}
We discuss our treatment of the CR electron energy spectrum in detail in \autoref{subsec:gamma}. We emphasize that including arbitrarily higher energies will not modify our results due to the power-law assumption also discussed in \autoref{subsec:gamma}. We discuss the spatial distribution and densities of CR electrons in detail in the next two sections.

We note here that we use cgs units in our synthetic observations and throughout our analysis.

\subsection{Model Eq: Cosmic Ray Equipartition}

In order to quantitatively investigate the accurate recovery of magnetic field measurements from radio intensities using equations that assume CR equipartition, we test our analysis throughout by imposing energy equipartition between the magnetic field and CRs. We consider equipartition at the scale of individual cells, which typically corresponds to 10 - 20 pc in the ISM of the simulated galaxy. We also test the impact on the synchrotron intensities of two other cases of equipartition: equiparting at the scale of 100 pc and equiparting at $\sim$10 - 20 pc followed by the diffusion of the CR electrons with a diffusion scale length of $\sim2$ kpc (further described below)

To compute equipartition at a given scale $\mathcal{L}$, we employ a similar approach to the `small scale turbulence computation' employed by \citet{Martin-Alvarez2022}, described in their Section~{2.3}. 
For each cell, and for each separate component of the magnetic field, we measure its coherent component as its average in a sphere of radius 
$0.5 \mathcal{L}_\text{scale}$. In this calculation we employ the full AMR grid reconstruction, subdividing cells intersecting the boundary of the sphere down to the highest level of refinement to capture their correct contribution down to our resolution limit. This per-scale computation is stored for each AMR cell, and the resulting magnetic fields $(B_x, B_y, B_z)_{\mathcal{L}}$ are employed to compute the magnetic energy used for establishing equipartition at a scale $\mathcal{L}$.

An alternative scenario to consider is that where equipartition is only held at the smallest scales of the galaxy, and the CRs distribution is then smoothed by their diffusion to larger scales. To approximate this scenario, we compute the density of energy for CRs equipartition employing a similar method to that above. First, we compute the equipartition CR energy density on a per-cell level. Then, we compute its diffusive re-distribution assuming it follows the behaviour of CRs under isotropic diffusion. We assume a diffusion coefficient $\kappa = 10^{28}~\cm^2 / \s$ \citep{Vollmer2020}, and diffusion on a timescale of 10 Myr. The resulting half-concentration radius is $\sim$$1\,\kpc$ (and $\mathcal{L}_\text{scale} \sim 2\,\kpc$), where the final energy for equipartition with CRs in each cell receives contributions from other cells up to a distance of $\sim$$8\,\kpc$. This energy is then stored for each cell, and employed by {\sc ramses2polaris} to establish diffused small-scale equipartition.

A critical consideration at this point is which component of CRs to consider in equipartition with the magnetic field energy. Throughout this work, we focus on the case of equipartition with the entire CR population (approximated by the protons, which dominate the CR energy budget) \citep{Beck2005RevisedObservations, BeckReview2015, Seta&Beck2019}. In \autoref{sec:Results_e}, we briefly consider the case of assumed equipartition between the CR electrons and the magnetic field, as employed by \citet{Reissl2019}. We present the methodologies used in both cases below.

\subsubsection{Model pEq: Cosmic Ray Proton Equipartition}
\label{sec:modelpEq}
We follow the arguments of \citet{Beck2005RevisedObservations} (hereafter BK05) to obtain an expression for the density of CR electrons given energy equipartition between the magnetic field and the CR protons. BK05 assume a small contribution from the CR electrons to this equipartition, modulated by the inverse of a constant $K_0$, which corresponds to the ratio of number densities of protons to electrons at an energy $E_0$:

\begin{equation}
    \label{eq:K0}
    \frac{n_{\rm p,0}}{n_{\rm e,0}} = K_0
\end{equation}

However, this contribution of electrons to the energy budget is small as $K_0$ is often assumed to be $\approx$ 100 (see \autoref{sec:EqB}), and we therefore consider the CR electrons negligible in our description of energy equipartition. We also assume heavier CRs than protons to be negligible in the energy integral (see direct measurements of the CR electron spectra separated by species in \citet{Cummings2016}). Such equipartition can be expressed quantitatively in terms of the energy densities as:
\begin{equation}
    \label{eq:eq_p}
    \epsilon_\text{CR} \approx \epsilon_\text{CR,p} = \epsilon_{B}
\end{equation}
where $\epsilon_\text{CR}$ is the energy density of all CRs and $\epsilon_\text{CR,p}$ the kinetic energy density of the CR protons. The energy density of the magnetic field $\epsilon_{B}$ is related to the magnitude of the field by:
\begin{equation}
    \label{eq:uB}
    \epsilon_{B} = \frac{B^2}{8 \pi}
\end{equation}
In order to determine the relation between the local number density of CR electrons and the local magnetic field that follows from \autoref{eq:eq_p}, we follow similar assumptions to BK05. We rewrite and discuss the assumptions here and refer to their Appendix for further details and explanations not expanded upon here. Above a minimum energy $E_{\rm min}$, we assume the number density of CRs (protons or electrons) per energy interval follows a power-law scaling with kinetic energy, resulting in the following total number of CRs (protons or electrons):
\begin{equation}
    \label{eq:nE}
    n_\text{CR} \equiv \int n_0 \left( \frac{E}{E_0} \right)^{-p} dE
\end{equation}
where $E_0$ is an energy normalization that is greater than $E_{\rm min}$. For CR protons, this minimum energy is the spectral break energy $E_{\rm b}$, or the energy at which the energy spectrum of protons steepens. We approximate this energy to be the rest energy of the proton, $E_{\rm p}$. Below this steepening at $E_{\rm b}$ (at lower energies), the spectrum is assumed to be flat. 
For the electrons, the $E_{\rm min}$ (or $E_{\rm b}$) is also assumed to be their rest mass energy ($E_{\rm e}$) in BK05.
We set $E_{\rm min}$ for electrons to be $4 E_{\rm e}$ (equivalent to the minimum Lorentz factor in \autoref{eq:gamma}) in order to parameterize deviations from a strict power-law description of the CR electron energy spectrum at low energy and discuss this choice in detail in \autoref{subsec:gamma}.

Integrating \autoref{eq:nE} between minimum and maximum kinetic energies of the form $E_{\rm min} = (\gamma_{\rm min}-1) mc^2$ and $E_{\rm max} = (\gamma_{\rm max}-1) mc^2$ (where the Lorentz factors characterizing the energy range are set in \autoref{eq:gamma}) and specifically considering the population of CR electrons yields the following expression:
\begin{equation}
    \label{eq:neE}
    n_{\rm CR,e} = \frac{n_{\rm e,0} E_0^p (E_{\rm min}^{1-p} - E_{\rm max}^{1-p})}{(p-1)}
\end{equation}
This normalization $n_{\rm e,0}$ is related to the normalization of the proton energy spectrum by Equation 7 in BK05, repeated in \autoref{eq:K0}. The normalization of the proton energy spectrum is, in turn, related to the energy density of CR protons through equation A13 of BK05. To compute A13, an integral over energy has been performed on $n(E)\, E\, dE$ above a nonrelativistic energy threshold $E_1$ (where $E_1 << E_{\rm p}$) assuming a power-law index $p > 2$:
\begin{equation}
\label{eq:ep}
    \epsilon_{\rm CR,p} = n_{\rm p,0} E_0^2\left( \frac{E_0}{E_{\rm p}} \right)^{p-2} \left( \frac{1}{2} + \frac{1}{p-2} \right)
\end{equation}
Combining \autoref{eq:K0} through \autoref{eq:ep} results in the following expression for how the total number density of CR electrons, assuming equipartition between the CR protons and the magnetic field, (Model pEq) scales with the local magnetic field $B$ and power-law index $p$:
\begin{equation}
    n_\text{CR,pEq}(x, y, z) = \frac{B^2(x, y, z) (p - 2) E_{\rm p}^{p-2}(E_\text{min}^{1-p} - E_\text{max}^{1-p})}{4 \pi K_0 p (p - 1)}
    \label{eq:nCRdensity_eq_p}
\end{equation}
Hereafter, we write $n_{\rm CR,e}$ as $n_{\rm CR}$, dropping $\rm e$ and $\rm p$ subscripts as we only discuss CR electron number densities throughout. The subscripts $\rm pEq$ and $\rm eEq$ are used in this section to distinguish the CR electron number density derived in our two distinct equipartition assumptions. The differences between our treatment and the treatment of BK05 are twofold: our neglect of the subdominant CR electrons in the energy equipartition with the magnetic field and our inclusion of the $E_{\text{min}}$ and $E_{\text{max}}$ term that explicitly accounts for the bounds of our energy integral. This term replaces a factor of $E_{\rm e}$ which would imply the power-law energy spectrum is integrated from $E_{\rm e}$ to an infinite maximum kinetic energy. We again refer the reader to \autoref{subsec:gamma} for a discussion on the choice of $E_{\rm min} \neq E_{\rm e}$.

\subsubsection{Model eEq: Cosmic Ray Electron Equipartition}
\label{sec:modeleEq}
We briefly consider the case of the CR energy equipartition with the magnetic field being dominated by the electrons in \autoref{sec:Results_e}. This general assumption has been applied for the pair plasmas of relativistic jets and for supernovae sources that are potentially dominated by electrons or positrons and emit synchrotron radiation \citep{Burbidge1956, Chevalier1998, Beck2005RevisedObservations, Duran2013}. It is not typically applied to the ISM of galaxies, however a version with electrons alone was employed in \citet{Reissl2019}. Though it might not be motivated for use in the context of the ISM, any ``true" form of equipartition is unknown to the observer. Therefore we also test the ramifications of this electron equipartition assumption, and whether it is capable of reproducing the observations. For this case, we follow the CR2 model found in \citet{Reissl2019}. The CR electron density scales with the local magnetic field $B$ following the relation:
\begin{equation}
    n_\text{CR,eEq}(x, y, z) = \frac{B^2(x, y, z) (p - 2)}{8 \pi E_\text{min} (p - 1)}
    \label{eq:nCRdensity_eq_e}
\end{equation}
This equation is derived by replacing the energy density of protons in \autoref{eq:eq_p} with the energy density of electrons and obtaining an analogous expression to \autoref{eq:ep} for $\epsilon_{\rm e}$ where the analogous integral has a lower bound of $E_{\rm min}$. 

\subsection{Model Exp: Prescribed Cosmic Ray Distribution}
\label{sec:modelB}
In order to probe how the equipartition assumption affects recovery of the magnetic field strengths when equipartition is not manifestly true, we prescribe a smooth distribution of CR electrons motivated by modeling of the Milky Way \citep{DrimmelSpergel2001, Page2007, Sun2008, Reissl2019}.
In this second model of the CR electron distribution, the number density $n_\text{CR}$ is set following the CR1 model in \citet{Reissl2019}:
\begin{equation}
    n_\text{CR,Exp}(r, z) = n_\text{CR,0} \exp\left(-\frac{r}{R_\text{gal}}\right) \left[\cosh\left(\frac{z}{h_\text{gal}}\right)\right]^{-2}
    \label{eq:nCRdensity}
\end{equation}
We consider two cases for the magnitude of the CR density at the center of the galaxy, which sets the amplitude of the density distribution (Model Exp1 and Exp2). In the first case (Model Exp1), $n_\text{CR,0} = 1.74\times10^{-4}\, \text{e}^{-}\, \cc$, corresponding to the number density of CR electrons in the center of the Milky Way. This is obtained by extrapolating the measured CR number density in the Solar neighborhood as determined in \citet{Sun2008} and \citet{Reissl2019}, based on GeV electron energy spectra from \citet{Barwick_1998} and \citet{Grimani}.
We select $R_\text{gal} = 8$ kpc and $h_\text{gal} = 1$ kpc to adjust the distribution to our simulated galaxy. This scale radius is also roughly consistent with the extent of the radiating CRs determined for the observed galaxy IC~342 with which we compare throughout \citep{GraveBeck1988}. While some observational studies have hinted at larger scale heights from synchrotron emission
\citep[where $h_{\rm gal} \geq h_{\rm syn}$;][]{Krause2014, Krause2018_CHANG-ES}, we note that once a minimum $h_\text{gal}$ is surpassed, the total intensity is approximately insensitive to the choice of vertical scale, and is instead modulated by the vertical scale of the magnetic field. This, and further details of the chosen scale height are discussed in Appendix B of \citet{tomography2024}. In our second case (Model Exp2), we set $n_\text{CR,0} = 1.74\times10^{-6}\, \text{e}^{-}\, \cc$, a factor of 10$^2$ lower than in Model Exp1. We keep all other values of the model prescribed CR distribution the same. We use this case to probe a lower regime of CR number densities. Such number densities are closer to the values yielded by the equipartition assumptions, but with the smoother distribution expected from a continued diffusion of CRs at galactic scales over the lifetime of a galaxy. We emphasize that these exponential models are intended as limiting cases: smooth distributions that do not assume or satisfy equipartition. Future work can expand such simple models by accounting for local variations such as enhancements due to star formation.

\subsection{Further Observational Effects}
For each combination of CR distribution model and power-law index $p$, we generate synthetic radio observations corresponding to Stokes parameter ($I$, $Q$, and $U$) maps at a wavelength of 6.2 cm and at a full-resolution approximately matching that of the simulation. We focus our study on the total intensity $I$ maps. These intensity maps are synthetically observed assuming a distance of 3.5 Mpc from the galaxy to the observer. This is to reflect the distance of the galaxy IC~342 analyzed in \citet{Beck2015}. To further match real observations, we account for beam convolution effects by convolving this intensity map with a 2D Gaussian kernel. We choose a kernel size corresponding to a full-width half-max of 25 arcseconds (corresponding to 0.417 kpc), the beam size of the radio intensity observations included in \citet{Beck2015}.

\section{Equipartition magnetic field strength measurement}
\label{sec:EqB}
As our goal in this work is to assess the validity and impact of equipartition assumptions in magnetic field strength inferences, we estimate the inferred magnetic field strength from our synthetic intensity observations analogously to the methods employed to analyze observations. This allows us to compare the resulting magnetic field measurements with the intrinsic values we obtain directly from the simulation.
We perform the conversion from intensity to magnetic field strength on each pixel of the convolved intensity map.
For a self-consistent treatment, and dependent on our choice of CR protons or CR electrons in energy equipartition with the magnetic field (Model pEq and eEq, respectively), we must make use of distinct equipartition magnetic field equations. Despite energy equipartition between magnetic energy and CR electrons alone not being a common assumption, in \autoref{sec:Results_e} we nevertheless test out its potential ramifications as the ``true" form if any of equipartition is unconstrained observationally. 

\subsection{CR Proton Equipartition (Model pEq)}
\label{sec:CRp}

We first discuss our primary focus of equipartition between the CR protons and magnetic field energy. The equation used predominantly in the literature to convert observed synchrotron intensities into equipartition magnetic field strength measurements \citep[e.g.,][]{Fletcher2011, Lacki2013TheLosses, BasuRoy2013, Krause2018_CHANG-ES} was presented in BK05. This BK05 equation is derived by combining the formula for synchrotron intensity with the CR number density equations derived using the treatment and assumptions outlined in \autoref{sec:modelpEq}, between \autoref{eq:K0} through \autoref{eq:ep} with two differences. The first difference is their inclusion of the subdominant CR electrons in the energy equipartition. To obtain the BK05 equation, one would have to modify \autoref{eq:eq_p} to be $\epsilon_\text{CR} \approx \epsilon_\text{CR,p}(1 + K_0^{-1}) = \epsilon_{B}$ where the inverse of the constant $K_0$ has been included to modulate the contribution of CR electrons and is defined in \autoref{eq:K0}. We emphasize that this altered assumption has a below percent level impact on the equipartition magnetic field estimate for the assumed $K_0$ value. The second difference between the CR density treatment used to obtain the BK05 equation and our treatment is that we explicitly integrate between the minimum and maximum kinetic energies defined by the Lorentz factors in \autoref{eq:gamma}. We discuss the motivations and implications of the minimum energy chosen in \autoref{subsec:gamma}. In short, this term accounts for our assumption on the shape of the CR electron energy spectrum. We rewrite the BK05 equation as:

\begin{equation}
\label{eq:Beck}
\begin{split}
B_\text{Eq, BK05}^{\frac{p+5}{2}} = \frac{4 {(2\pi)}^{\frac{p+3}{2}} p (p+1) I_{\lambda} (K_0+1) E_{\rm e}^{\frac{3p-1}{2}}}{3^{\frac{p}{2}} (p-2) \, \lambda^{\frac{p-1}{2}} L \, \text{cos}(i)^{\frac{p+1}{2}} e^{\frac{p+5}{2}} E_{\rm p}^{p-2} }\\
\times \frac{1}{\Gamma \left( \frac{3p - 1}{12} \right) \Gamma \left( \frac{3p + 19}{12} \right)}
\end{split}
\end{equation}

where $I_{\lambda}$ is the intensity measured at the given wavelength, $i$ is the inclination angle of the galaxy, $E_{\rm p}$ is the rest energy of the proton, $E_{\rm e}$ is the rest energy of the electron, $L$ is the approximate vertical thickness of the emitting galaxy disk, and $K_0$ is the ratio of proton to electron number densities at an energy $E_0$ (see \autoref{eq:K0}). We note that the power-law index $p$ is related to the synchrotron spectral index by $\alpha = (p-1)/2$. We hereafter refer to this assumption of equipartition used frequently in the literature as ``BK05 equipartition".

We modify this equation to reflect equipartition between only the CR protons and the magnetic field, and to reflect the energy bound of the integral with $E_{\text{min}}$ that we use as a suppression term to model deviations from the BK05 full power-law description of CR electrons (see \autoref{subsec:gamma}). We do this by combining the modified equation we derived for CR electron number density (\autoref{eq:nCRdensity_eq_p}) with the synchrotron intensity formula found in equation A15 of BK05, which is largely identical to earlier synchrotron formulae found in works such as \citet{Pacholczyk1970}. This formula relates the intensity at a particular frequency to the number density of CR electrons, the magnetic field strength perpendicular to the line of sight, the frequency of observation, and the power-law index. Combining the formula with our CR electron density equation results in the following equation for the equipartition magnetic field strength:

\begin{equation}
\label{eq:Beck_p}
\begin{split}
B_\text{pEq}^{\frac{p+5}{2}} = E_\text{min}^{p-1} \frac{4 {(2\pi)}^{\frac{p+3}{2}} p (p+1) I_{\lambda} K_0 E_{\rm e}^{\frac{p+1}{2}}}{3^{\frac{p}{2}} (p-2) \, \lambda^{\frac{p-1}{2}} L \, \text{cos}(i)^{\frac{p+1}{2}} e^{\frac{p+5}{2}} E_{\rm p}^{2-p}}\\
\times \frac{1}{\Gamma \left( \frac{3p - 1}{12} \right) \Gamma \left( \frac{3p + 19}{12} \right)}
\end{split}
\end{equation}
To continue our measurement of magnetic field strength maps using \autoref{eq:Beck_p}, we must set the values of the following constants. We set $\lambda$ = 6.2 cm, the wavelength corresponding to our synthetic intensity observation. We set inclination $i$ = 0 since in our analysis we view and measure the galaxy face-on. Note that small inclinations have a negligible impact on the overall synthetic intensity maps \citep{tomography2024}. We assume a $K_0$ of 100, motivated by the ratios of CR protons and electrons estimated for supernovae remnant shock fronts as well as by observed Galactic CR data \citep{Bell1978_2}. This number has been assumed in other works including but not limited to \citet{Bell2004, Beck2005RevisedObservations, BasuRoy2013, BeckReview2015, Beck2015}. We assume a full-thickness of the disk contributing to the intensity of $L$ = 1 kpc based on the thickness of the gas, stellar component, and magnetic field of this simulated galaxy \citep{Martin-Alvarez2020, tomography2024}. This value also matches the assumed thickness for similar spiral galaxies such as IC~342 \citep{Beck2015}.

\subsection{Cosmic Ray Electron Equipartition (Model eEq)}
\label{sec:CRe}
For our secondary case of equipartition only between the radiating CR electrons and the magnetic field (\autoref{sec:Results_e}), we begin with the equation presented in \citet{Reissl2019} that relates a magnetic field to the synchrotron emission produced when given a CR density. This equation governs how {\sc ramses2polaris} and {\sc polaris} calculate intensities from the intrinsic magnetic fields in the simulation. It is consistent up to the factor of $E_{\rm min}^{1-p}$ (discussed in \autoref{subsec:gamma}) with the synchrotron emission formulae presented in \citet{BlumenthalGould1970}, \citet{Leung2011}, and \citet{Pandya2016}. The coefficient of total emission is:

\begin{equation}
\label{eq:polaris_j}
\begin{split}
    j_\text{I} (\lambda) =\frac{E_\text{min}^{1-p}}{(E_\text{min}^{1-p} - E_\text{max}^{1-p})}  \frac{n_\text{CR} e^2 3^{\frac{p}{2}} (p-1) \,\text{sin}(\theta)}{2 (p+1) \lambda_{\rm c}} \\
    \times  \left( \frac{\lambda_{\rm c}}{\lambda \,\text{sin}(\theta)} \right)^{-\frac{p-1}{2}} \Gamma \left( \frac{3p - 1}{12} \right) \Gamma \left( \frac{3p + 19}{12} \right)
\end{split}
\end{equation}

where we have rewritten the equations to be explicitly expressed in terms of the CR electron kinetic energies. Here, $\theta$ is the angle between the magnetic field and the observer's line-of-sight, and $\lambda_{\rm c}$ is the critical wavelength corresponding to the critical frequency $\nu_{\rm c}$. It is defined as
\begin{equation}
    \lambda_{\rm c} = \frac{2\pi m_{\rm e} c^2}{e B}
\end{equation}
We combine this equation with the electron equipartition CR number density found in \autoref{eq:nCRdensity_eq_e} to derive the electron equipartition magnetic field in the plane of the disk galaxy at inclination $i$:

\begin{equation}
\label{eq:polaris_B}
\begin{split}
B_\text{eEq}^{\frac{p+5}{2}} = \frac{8 {(2\pi)}^{\frac{p+3}{2}} (p+1) I_{\lambda} E_{\rm e}^{\frac{p+3}{2}}}{3^{\frac{p}{2}} (p-2) \, \lambda^{\frac{p-1}{2}}\,L \, \text{cos}(i)^{\frac{p+1}{2}} e^{\frac{p+5}{2}}}\\
\times \frac{(E_\text{min}^{1-p} - E_\text{max}^{1-p})}{E_\text{min}^{-p}} \frac{1}{\Gamma \left( \frac{3p - 1}{12} \right) \Gamma \left( \frac{3p + 19}{12} \right)}
\end{split}
\end{equation}
An important assumption made in all the observational analyses that follow these considerations is that $j_{\lambda}$ can be approximated as $\frac{I_{\lambda}}{L}$ where $L$ is the approximate thickness of the galaxy disk that dominates the emission. As with CR proton equipartition, we set $L$ = 1 kpc, $\lambda$ = 6.2 cm, and $i = 0$ for the face-on galaxy.

\section{Results}
\label{sec:results}
\subsection{Intrinsic Magnetic Field Strength Profiles}
\label{sec:resultsBint}
In order to probe the validity of equipartition and its recovery of the intrinsic magnetic field profile, it is first instructive to consider that intrinsic profile and the information it summarizes about the galaxy.
Depending on how the profiling is performed, radial profiles can probe different regions and phases of the galaxy. This is determined by the weighting employed for the resolution elements, which all feature different masses and volumes.

\begin{figure}[t]
\includegraphics[width=\columnwidth]{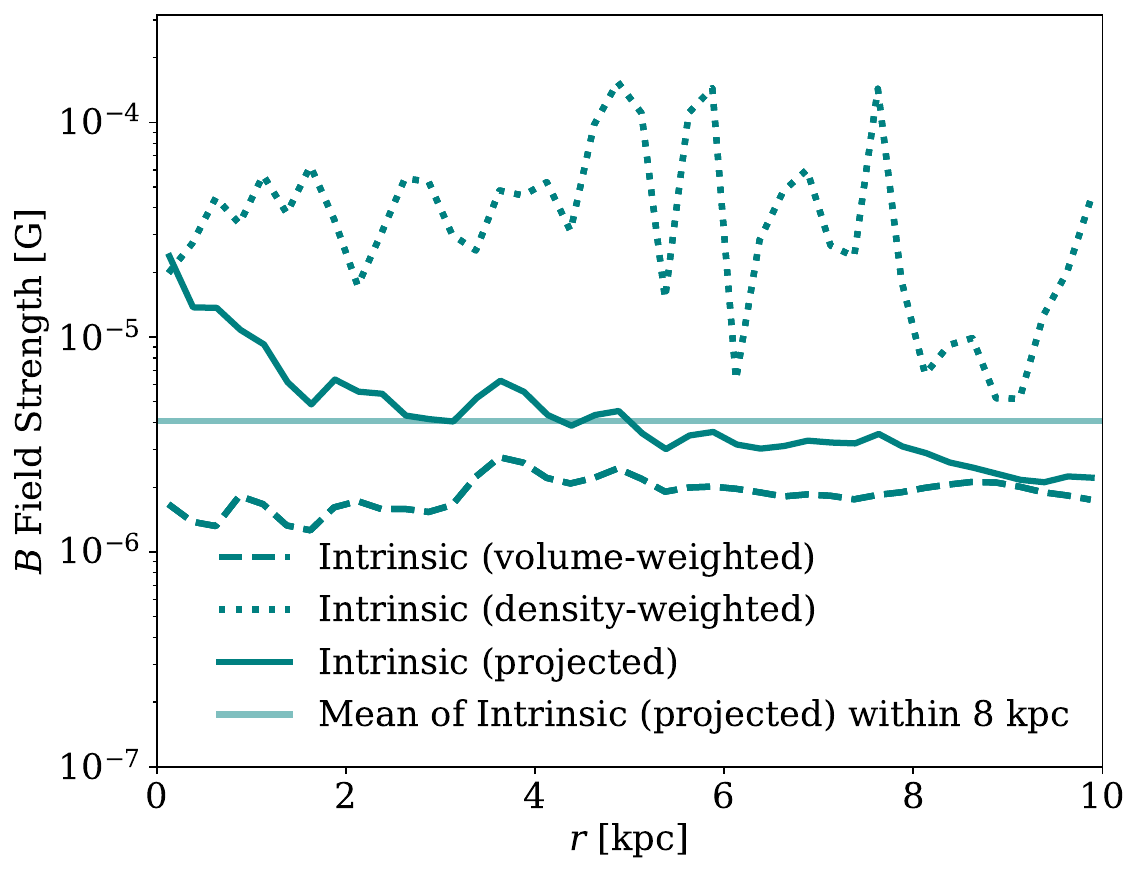}
    \caption{Intrinsic magnetic field strength profiles of the \nut~galaxy measured employing three different weighting schemes. The dashed (volume-weighted) and dotted (density-weighted) profiles are azimuthally averaged within a cylindrical region of 1 kpc thickness. We measure a density-weighted along the line-of-sight (face-on) profile creating a projected map followed by area-weighting in the radial profiling (solid line). We also include a horizontal line marking the average intrinsic magnetic field strength within 8 kpc, a value used to compute the theoretical backgrounds of our subsequent figures (see \autoref{fig:intensity}, \autoref{fig:square_comparison}, and \autoref{fig:square_e}). The density and volume weightings probe different regions of the galaxy. The projected profile matches the density-weight at the low radius limit and the volume-weight at high radius limit.}
    \label{fig:profiles_intrinsic}
\end{figure}

\begin{figure*}[ht]
\centering
\includegraphics[width=7in]{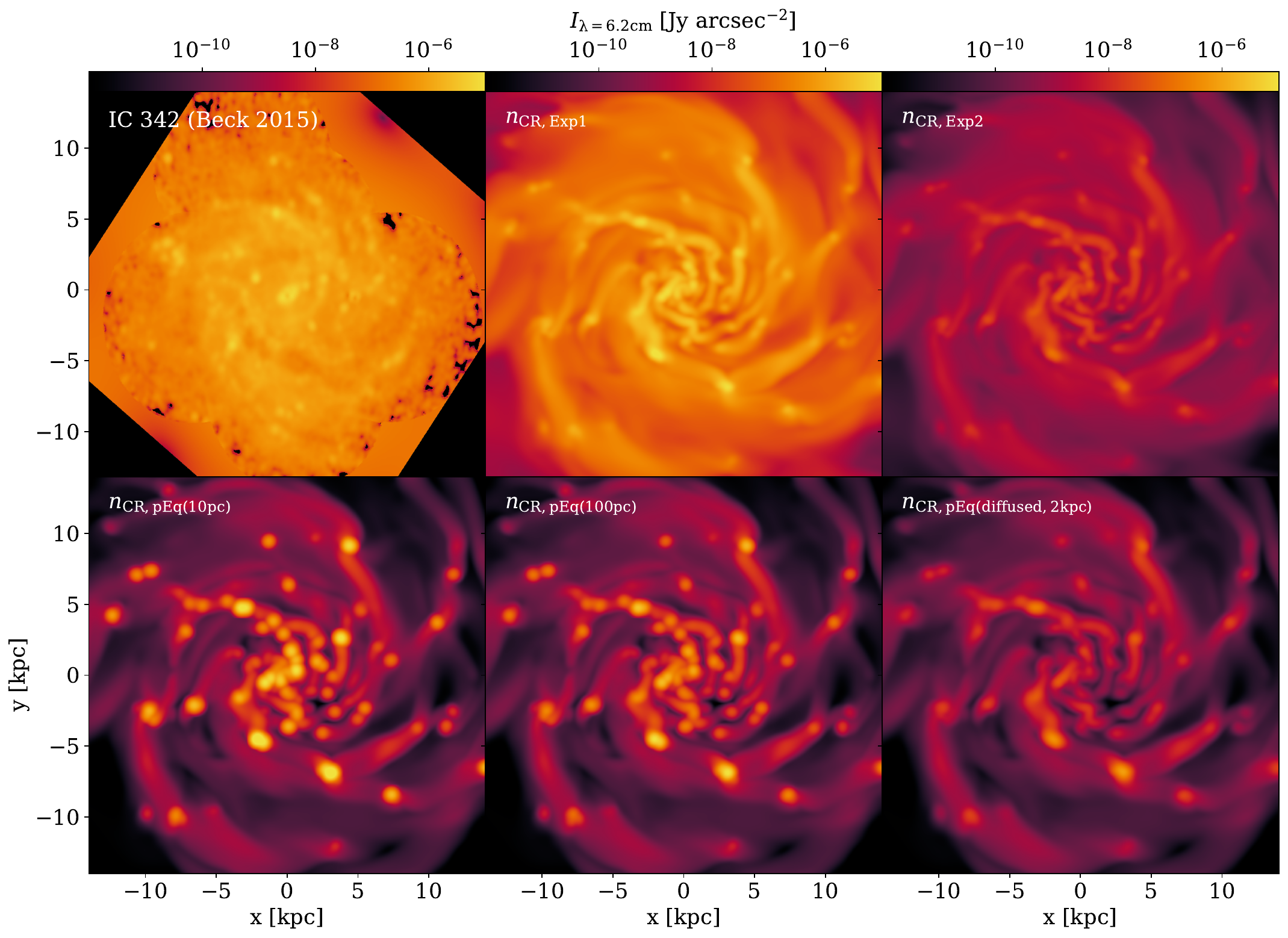}
    \caption{Synthetic observations of the galaxy's synchrotron intensity assuming power-law index $p$ = 2.9 for our five $n_{\rm CR}$ models. These correspond to the two cases of prescribed exponential CR distribution "Model Exp" in the top row, and three cases of equipartition CR distribution "Model pEq" at different scales in the bottom row. See \autoref{sec:synth_obs} for details on the synthetic observation. Model Exp cases result in smoother maps (with different amplitudes of intensity for the two cases), while the Model pEq cases display large dynamic ranges of low to high intensities across different spatial regions. Included in the top leftmost panel is the observation of IC~342 by \citet{Beck2015} for qualitative comparison. Note that the observed total intensities for IC~342 include a small contribution from thermal intensities (see \citealt{Beck2015}). We highlight the striking visual resemblance between IC~342 and the Model Exp1 case.}
    \label{fig:I_maps}
\end{figure*}

We explore three alternative weightings in \autoref{fig:profiles_intrinsic}, each summarizing different aspects of the magnetic field distribution across the galaxy, with important interrelations between them. The first two are the standard density-weighted (using total gas density) and volume-weighted spherically-averaged profiles of the magnetic field, shown with dotted and dashed curves respectively. These two profiles were azimuthally averaged within concentric cylindrical annuli of 1 kpc vertical thickness to focus our measurement on the disk of the galaxy, where we align the $z$ component with the angular momentum vector of the disk. We note that volume-weighted profiles shift to lower magnetic field strength values with increasing vertical thickness included, while the density-weighted profile remains dominated by the high-density disk. The 1 kpc thickness is also consistent with the disk thickness assumed in our inference of the magnetic field from our synthetic observations (see \autoref{sec:CRp}), and encompasses well the gas thick disk \citep{Martin-Alvarez2020}. The third profile (solid curve) is the surface area-weighted azimuthal average of the magnetic field strength, measured in density-weighted projections into the 2D disk plane. This procedure is meant to reflect how the intensity profiles would be created observationally. 

As expected, the density-weighted profile (\autoref{fig:profiles_intrinsic}, dotted line) is noticeably higher and more irregular than the other two profiles. It typically features field strengths of $\sim$$4\times 10^{-5}$ G, and reaches peak values of $10^{-4}$ G. The volume-weighted profile (dashed line) is relatively flat at all radii and is consistently weaker than the projected profile (solid line) at all radii. It typically features values of a few $10^{-6}$ G, with important contributions from the volume outside of the dense galactic disk. Note that the projected profile matches the density-weighted profile at the smallest radii where it is dominated by its line-of-sight density weighting and little area, allowing only for small surface variations. Conversely, at large radii, the projected profile matches the volume-weighted profile. This is due to the density variations being larger, with a vast dynamic range within each projected annuli. 
\begin{figure}[t]
\includegraphics[width=\columnwidth]{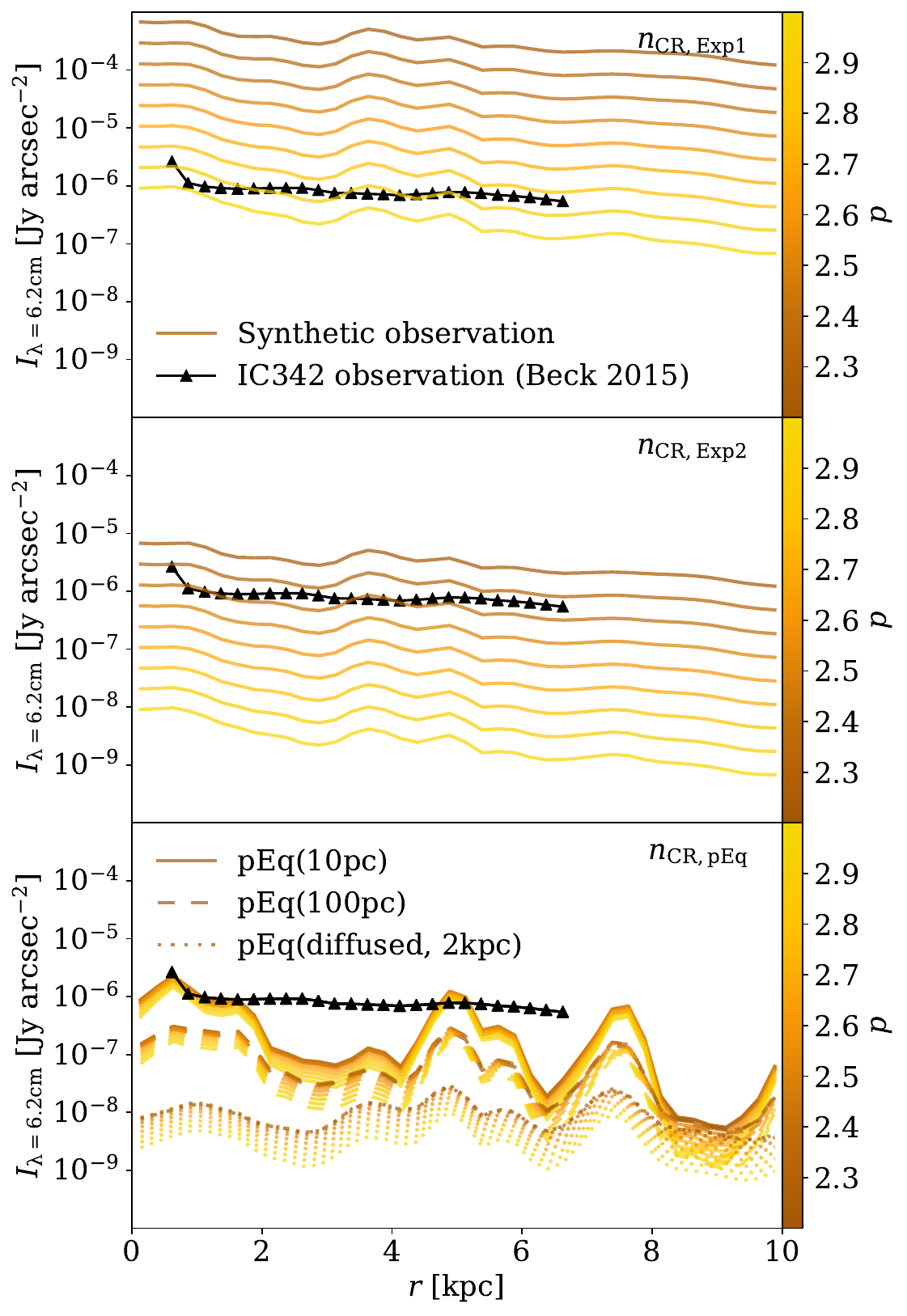}
    \caption{Radial synchrotron intensity profiles of the synthetically observed face-on galaxy shown for the two models of $n_{\rm CR}$: Model Exp and Model pEq (two cases of prescribed exponential CR distribution in the top two panels and three scales of equipartition CR distribution in the third panel). Included for comparison in black triangles is the radial intensity profile of IC~342 (see \citet{Beck2015}). Our radial intensity profiles are averaged in rings of 0.25 kpc, selected to be comparable to those employed by \citet{Beck2015}. The prescribed exponential CR distributions match the smoothness and radial flatness of the IC~342 observation. Equipartition on these scales leads to significantly less smooth profiles of lower intensity, which are dissimilar to observed galaxy intensities.}
    \label{fig:profiles_I}
\end{figure}
The projected profile provides the most faithful interpretable approximation to what is probed by the ``observed'' synchrotron emission and we consider it as the ``intrinsic" magnetic field strength profile moving forward.

\subsection{Synthetically Observed Intensities}
\label{sec:resultsI}

Combining the intrinsic magnetic field from the simulation with a model for the distribution of CR electrons, we generate maps of synthetically observed synchrotron intensity including beam convolution effects (see \autoref{sec:synth_obs}). We examine the intensity produced by five cases of CR electron distribution. For each case and for a range of power-law indices $p$ describing the CR energy spectrum, we can produce synchrotron intensity maps at 6.2 cm (see \autoref{sec:synth_obs}). We show the resulting intensity maps obtained with $p = 2.9$ for the five cases in \autoref{fig:I_maps} along with the observations of a representative spiral galaxy, IC~342 by \citet{Beck2015}, for qualitative comparison (upper leftmost panel). We compare to this galaxy throughout as it is nearby, nearly face-on, and is well-studied in the radio with high angular resolution \citep{Beck2015}. Note that for IC~342 we show the observed total intensity, which includes a small contribution from the thermal intensity ($<15\%$). The first two CR electron cases (see the rightmost upper two panels of \autoref{fig:I_maps}) are the Model Exp cases of high and low prescribed CR electron distribution discussed in \autoref{sec:modelB}. The last three cases (see the lower three panels of \autoref{fig:I_maps}) are three versions of Model pEq, imposed equipartition between the CR protons and the magnetic field which results in the CR electron distribution derived in \autoref{sec:modelpEq}. The three equipartition cases include equipartition at the scale of 10 pc, equipartition at the scale of 100 pc, and equipartition at the scale of 10 pc with subsequent diffusion of the CRs out to a diffusion scale of $\sim$2 kpc (see \autoref{sec:modelpEq}). The imposed equipartition cases result in clumpier intensity maps with regions of drastically lower intensity compared to the smoothness of the two prescribed CR distribution cases. The diffusion case has smoother clumps of intensity, but even the large kernel of diffusion does not reproduce the smoothness seen in the prescribed CR distribution cases and the IC~342 data. 

In \autoref{fig:profiles_I}, we present the radial intensity profiles for our five CR electron cases, varying the power-law index between $p = 2.2$ and $p = 3.0$. The two cases of Milky Way-like prescribed CR electron distribution (upper two panels) result in similarly-shaped smooth and shallow curves, shifted to significantly lower intensities for higher power-law index $p$ values. This extreme variation arises from imposing a fixed $n_{\rm CR}$ distribution and varying $p$ in our synthetic intensity observation alone.  For shallower, power-law slopes (i.e., lower $p$), the CR energy spectrum features a higher proportion of higher energy CR electrons that provide larger contributions to the intensity. In the last three cases of imposed equipartition (bottom panel), the profiles feature only small variations with $p$. This is because $p$ affects both the $n_{\rm CR}$ distribution as well as the synthetic emissivities. Notably, the three equipartition intensity profiles are overall jagged with large troughs of low intensity. The case with equipartition at 100~pc has slightly less severe deviations from their average, now approximately centered around the lower end of the 10~pc case. The case of diffusion out to 2~kpc results in smoother radial profile shapes but further suppressed intensities at all radii. We also tested equipartition on scales of 1~kpc, which reproduced the behavior of the other cases and resulted in even lower intensities (not shown) similar to the diffusion case. In other words, the price for smoothing the distributions of CR electrons under equipartition is even lower intensities.

In all three panels, we overplot the observed radial nonthermal intensity profile of the galaxy IC~342 as presented in \citet{Beck2015}. We include the nonthermal intensity profile out to only 7 kpc, beyond which the spectral index for the separation of thermal and nonthermal components is not well determined \citep{Beck2015}. We note that IC~342 is not entirely face-on and has an inclination angle of $i = 31\deg$ which is accounted for in the profile. The observed profile of IC~342 is well matched both in terms of shape and intensity values by Model Exp2 ($n_0 = 1.74 \times 10^{-6}\, {\rm cm}^{-3}$ and $p = 2.4$), and especially by Model Exp1 ($n_0 = 1.74 \times 10^{-4}\, {\rm cm}^{-3}$ and $p = 2.9$), illustrating the degeneracy between $n_{\rm CR}$ and $p$ in the resulting intensities. The Model Exp1 ($n_0$, $p$) values provide a more realistic match to IC~342, as they reproduce the local Milky Way CR electron properties (see \autoref{sec:modelB}). For the equipartition cases, the observed profile is in noticeable disagreement with the steep intensity troughs and overall low intensities seen in the three imposed cases studied. This suggests that true equipartition at scales of 10~pc, 100~pc, 1~kpc (not shown) and even with diffusion out to 2~pc, is not able to produce the smooth intensity distributions of observed galactic synchrotron profiles. This result is consistent with the findings of \citet{Stepanov2014}, \citet{Seta2018}, and \citet{Seta&Beck2019}, indicating that equipartition does not hold on scales below 100~pc and even 1~kpc. It also reinforces the claim by \citet{Stepanov2014} that equipartition leads to fluctuations in the synchrotron intensity far exceeding what is observed, suggesting smoother and less strongly varying CR density distributions in such galaxies. We note that depending on the details of the treatment of the CR electron energy spectrum in the equipartition assumption, the intensities have some freedom to move to higher values (closer to the IC~342 observation). However, the favored expectation is that in reality imposing equipartition would result in even lower intensities that are even more discrepant with observations. This is discussed further in \autoref{subsec:gamma}.

\begin{figure}[t]
\centering
\includegraphics[width=.5\textwidth]{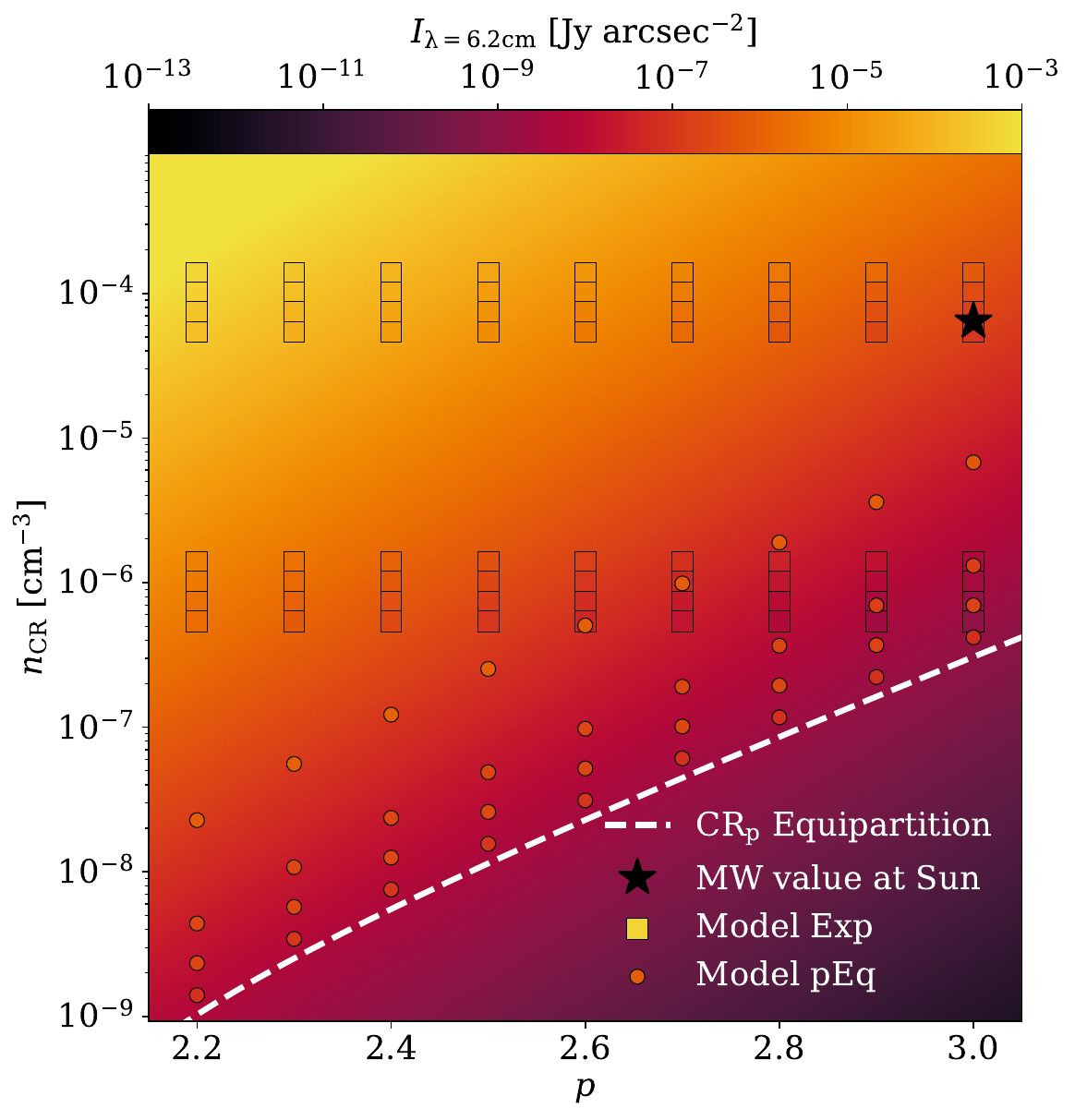}
\caption{Intensity variation as a function of power-law index $p$, $n_{\rm CR}$, and radius (4-point series). We show both the prescribed $n_{\rm CR}$ (squares) and equipartition $n_{\rm CR}$ (circles) models. We color the background according to the theoretical prediction for the synchrotron intensity assuming the average magnetic field of the galaxy (see text). The white line is the theoretical expectation for the equipartition $n_{\rm CR}$ as a function of $p$ for that same mean magnetic field strength. Finally, we include a black star indicating the estimated local Milky Way $n_{\rm CR}$ and $p$ as a reference (see text). Large regions of the $p$ and $n_{\rm CR}$ space, particularly that occupied by equipartition (circles and white line), yield extremely low intensities, incompatible with those observed in galaxies.}
\label{fig:intensity}
\end{figure}

In \autoref{fig:intensity}, we summarize the intensity information contained in the profiles as functions of the CR electron density $n_{\text{CR}}$ and power-law index $p$. 
Each vertical series of squares in \autoref{fig:intensity} displays the magnitude of the galaxy's intensity profiles for the two Model Exp cases. The four vertical squares correspond to four wide radial bins between 0 and 10 kpc within which the intensity profile is averaged, with the color of each square corresponding to the average intensity within that radial bin. The $n_\text{CR}$ value for each radial bin (i.e., each square) is determined by averaging its values in a similar manner. As $n_\text{CR}$ decreases with radius, the top square of a vertical series corresponds to the innermost radial bin, and the bottom square to the outermost. The vertical series of circles displays the same information but for Model pEq(10pc), where we impose CR equipartition at a scale of 10 pc. To obtain the values of $n_\text{CR}$ in radial bins for Model pEq(10pc), we generate maps of the resulting $n_\text{CR}$ from equipartition, compute their radial profiles, and then average their number density using the same radial bins as described above. Throughout the rest of this work, we focus on the Model pEq(10pc), which leads to the closest intensities to the observations. We stress that the qualitative differences between the equipartition cases do not meaningfully impact our later results. 

The smooth background colors in \autoref{fig:intensity} display the theoretical prediction for the intensity in this parameter space by assuming a vertical ISM slab with uniform magnetic field perpendicular to the line-of-sight, and calculating \autoref{eq:polaris_j} for each point in $n_\text{CR}$ and $p$ space. The uniform magnetic field employed for these predictions ($\sim$$4 \mu$G), corresponding to the average intrinsic projected magnetic field of the simulated galaxy within 8 kpc ($B$, delineated in \autoref{fig:profiles_intrinsic} as the horizontal solid line). The near perfect agreement between the theoretical background and the square points from the simulated galaxy intensity is due to the smooth and shallowly decreasing nature of the intensity profiles in the cases of prescribed equipartition (Model Exp1 and Model Exp2) (see \autoref{fig:profiles_I}). We include a white dashed curve showing the theoretical expectation for the unique $n_\text{CR}$ values defined by CR equipartition in \autoref{eq:nCRdensity_eq_p} as a function of $p$, and a fixed intrinsic $B$ (set to the same average projected $B$ within 8 kpc). 

The black star in \autoref{fig:intensity} indicates estimates of the Milky Way $p$ and $n_\text{CR}$ in the Solar vicinity \citep{Sun2008}. Note that the Milky Way might possess an intrinsic magnetic field slightly different from the value we employ here from our simulated galaxy, and is included in the figure to provide observational context for the expected number density and power-law index of CRs in the ISM of galaxies. The Milky Way total magnetic field strength unperturbed by the draping around the heliosphere has been modeled to be around 2.9 $\mu$G using a combination of MHD simulations of the solar wind and local ISM and observations of NASA's Interstellar Boundary Explorer, and matched to Voyager data \citep{Zirnstein2016}. This magnetic field value, if used to compute the background and white dashed line expected by equipartition, would actually shift them slightly such that equipartition corresponds to even lower intensities.

Both the theoretical background and our synthetic intensity measurements (\autoref{fig:intensity}) demonstrate how for a given intrinsic magnetic field, there are sizeable regions in ($n_\text{CR}$, $p$) parameter space that are effectively ruled out due to their drastic deviation from the observed radio intensities in comparable systems. E.g., are the 6.2 cm nonthermal intensities of IC~342 at $\sim$$10^{-6}$ Jy/arcsec$^{-2}$  \citep{Beck2015}, M51 at $\sim$$10^{-4}$ Jy/arcsec$^{-2}$ \citep{Fletcher2011}, and NGC 6946 at $\sim$$10^{-3}$ Jy/arcsec$^{-2}$ \citep{Beck2007}. Even for our reasonably strong magnetic field at $\sim$4 $\mu$G (see \autoref{fig:profiles_intrinsic} and \citet{Martin-Alvarez2020}), the intensities predicted by the imposed equipartition case (circular points) are at least an order of magnitude lower than those observed in similar galaxies. We do not display points corresponding to the 100~pc, 1~kpc and diffusion out to 2~kpc equipartition cases, for the sake of readability, but we note their intensities are even lower than the 10~pc case. As the intrinsic magnetic field is reasonably large, these results suggest that CR equipartition produces CR densities that are not sufficient to generate the radio intensities observed.

The disagreement between observations and the intensities resulting from imposing equipartition as well as the shape of their radial profiles suggest that this equipartition scenario (at a range of scales) is disfavored by both theoretical predictions and synthetic observations of our simulations. This finding complements the claims of \citet{Seta2018} and \citet{Seta&Beck2019} which assert that CR energy densities and magnetic field energy densities are statistically independent and uncorrelated at scales smaller than approximately 100 pc, making it unlikely for equipartition to be valid at smaller scales. We have shown further that even at scales of 100~pc and above, or even in a diffusion out to 2~kpc scenario, equipartition is unable to match observed intensities. As a result, the assumption of CR energy equipartition to compute magnetic field strengths from the observed intensities is likely inaccurate, and could lead to over or underestimation of the magnetic field strengths. We review this in the next section.

\begin{figure*}[t!]
\centering
\includegraphics[width=.9\textwidth]{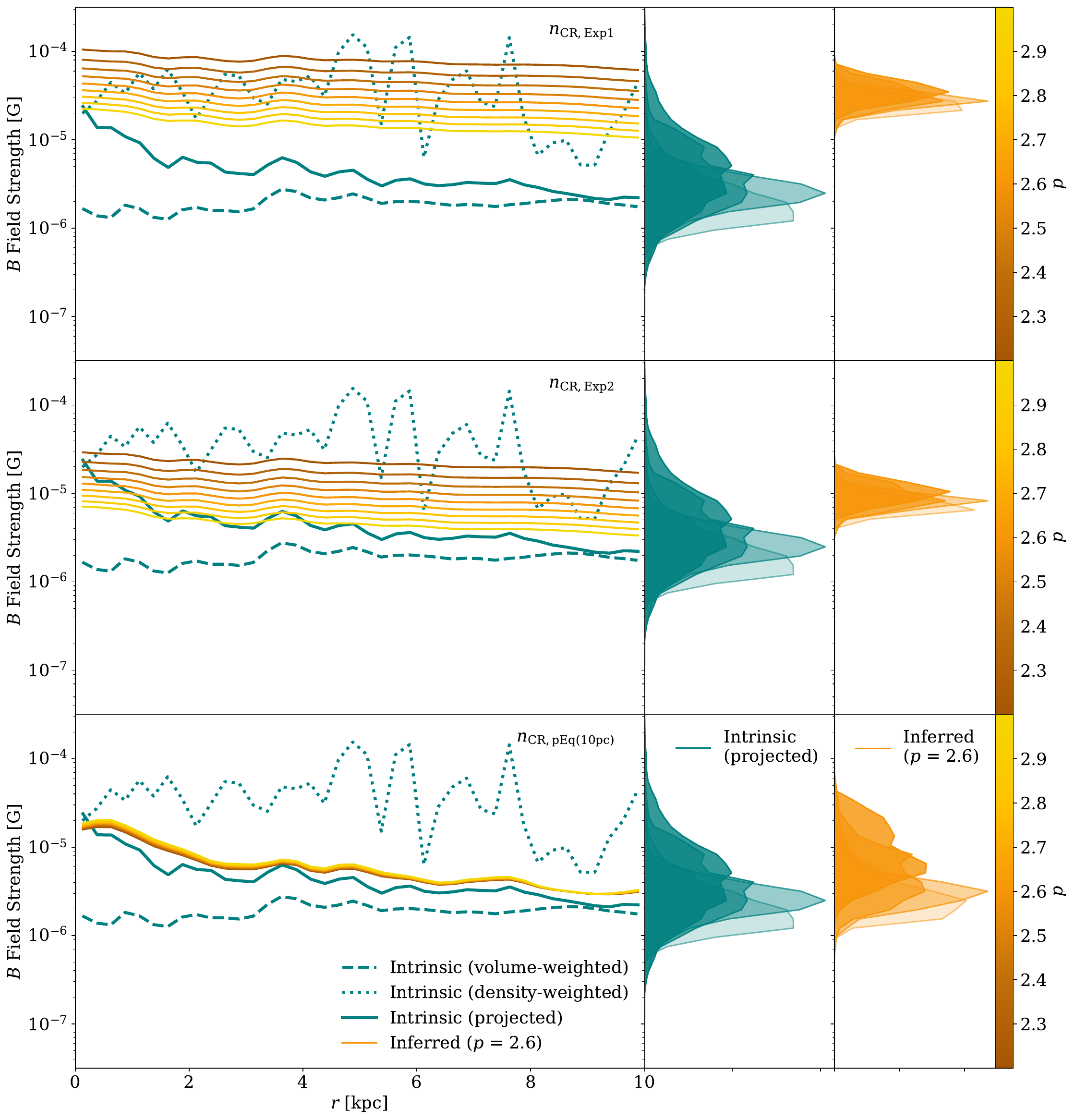}
    \caption{Comparison between the intrinsic magnetic field profiles (teal lines) and those inferred from synthetically observed radio synchrotron maps. The inferred magnetic field strengths is computed following the procedure used observationally with the assumption of CR and magnetic energy equipartition. We show the inferred magnetic fields for a range of power-law indices $p$, shown as varying orange shades. This comparison is done for three of our models for $n_{\rm CR}$ (two cases of prescribed smooth CR distribution and one case of equipartition CR distribution). In the second column, we show the probability distribution functions of the intrinsic projected magnetic field strength pixels, separated into large radial bins of 2.5 kpc. In the third column, we show the same PDF for the $p$ = 2.6 magnetic field strength in the same large radial bins. When equipartition is imposed to for the underlying distribution of $n_{\rm CR}$ in the galaxy, the inferred magnetic field strength profile recovered using the equipartition assumption provides a reasonable match to the intrinsic magnetic field. However, for our $n_{\rm CR}$ Exp models that produce realistic intensities, the inferred magnetic field deviates considerably from the intrinsic profile that should be recovered, with the deviation being a strong function of $p$.}
    \label{fig:profiles_B}
\end{figure*}

\subsection{Intrinsic vs Inferred Magnetic Field}
\label{sec:infvint}
In this section, we examine the implications of applying the equipartition assumptions when the observed galaxy does not fulfill such an assumption. Specifically, we focus on its impact on the estimation of magnetic field strengths (perpendicular to the line-of-sight), studied as a function of the intrinsic CR distribution, the CR power-law index $p$, and the different ISM environments of the galaxy.
Using the synthetically observed intensity maps discussed in \autoref{sec:resultsI} along with \autoref{eq:Beck_p}, we obtain inferred magnetic field strength maps for each case of prescribed CR distribution and imposed equipartition CR distribution, varying the power-law index $p$ across our studied range. \autoref{fig:profiles_B} displays the azimuthally-averaged radial profiles obtained from these inferred magnetic field strength maps, alongside the intrinsic magnetic field profiles discussed in \autoref{sec:resultsBint}. 

The 10~pc imposed equipartition case (third row) serves as a best-case scenario test for reconstructing the intrinsic magnetic field, as it explores the situation when the equipartition assumption is {\bf exactly true}. As expected, reversing the assumption employed to generate the emission provides a reasonable recovery of the intrinsic magnetic field. The accuracy of this recovery is made more evident in the second and third panels of the third row, which show the probability distribution functions of the magnetic field strength pixel values that contribute to the profile in four large radial bins. These distributions are almost entirely overlapping for each corresponding radial bin. The slight overestimation by the equipartition magnetic field strength compared to the intrinsic projected field strength is expected due to the non-linear relationship between synchrotron intensity $I_{\lambda}$ and $B_{\rm tot,\perp}$, modulated by line-of-sight fluctuations in $B_{\rm tot,\perp}$ \citep{Beck2003, Stepanov2014, Seta&Beck2019, BeckReview2015}. Another assumption that contributes to the lack of exact recovery is the assumption of the thickness of the emitting region $L$, assumed to be a constant 1~kpc throughout. This nevertheless close recovery indicates that the magnetic field probed by synchrotron emission can be physically interpreted as the line-of-sight density-weighted magnetic field. This result holds despite, and independently of, the demonstrated failure of the equipartition assumption in generating realistic synchrotron intensities.

The upper rows of \autoref{fig:profiles_B} display our intrinsic CR configurations following a smooth distribution of the exponential form (\autoref{eq:nCRdensity}), capable of matching the observed intensity profiles for realistic $n_{\rm CR}$ values. However, such CR distributions face significant overestimation of the magnetic field strength with respect to the intrinsic projected profile due to incorrectly applying the equipartition assumption (top and central rows of \autoref{fig:profiles_B}). This overestimation strongly depends on the assumed $p$. The rightmost panels of these rows show an inferred magnetic field with considerably narrow distributions in the four radial bins that does not match the spread of the intrinsic magnetic field (central column). Synchrotron emission is the combined effect of the CR electron density, CR spectral energy distribution, and the magnetic field configuration. The misestimation found when the equipartition assumption is applied to infer the magnetic field is driven by the fact that the same emission can be generated by multiple combinations of magnetic field strengths and CR electron densities. The equipartition assumption simply fixes them to a specific, but not necessarily true, combination.
Assuming equipartition when it is not valid results, not in a simple offset in the magnetic field, but in a different profile shape altogether. Equipartition will always result in a magnetic field strength profile proportional to a specific power of the intensity profile for a constant assumed power-law index (see \autoref{eq:eq_p}). Therefore depending on the details of the CRs distribution, this assumption may limit the ability of observations to recover important features of the magnetic field across different parts of galaxies.

We emphasize here that the equipartition assumption used throughout this work is an observationally and theoretically motivated adjustment to the BK05 equipartition formula. Factors of overestimation shown are therefore with respect to our motivated equipartition, and would require division by a constant factor of 1.5-2 to recover the over- and underestimation expected from the BK05 equipartition assumption used frequently in the literature. We discuss the motivations and consequences of our treatment in detail in \autoref{subsec:gamma}, along with further detail on this constant factor. The observationally inferred total magnetic field strength in IC~342 using the BK05 equipartition assumption is 15~$\mu$G, averaged within the inner $\sim$3~kpc, and 13~$\mu$G within $\sim$6~kpc \citep{Beck2015}. Our simulated galaxy with a mean magnetic field strength within 8 kpc of $\sim$4 $\mu$G, in combination with the Model Exp1 CR densities and $p=2.9$, is able to match those inferred BK05 equipartition magnetic field strengths (accounting for the factor of $\sim$2 noted as necessary for such a comparison above). This is consistent with the CR model that generates an intensity profile matching IC~342 in \autoref{fig:profiles_I}. This suggests that the factor of 2-3 difference between the inferred magnetic field strength of IC~342 and the intrinsic magnetic field strengths in our simulation could be the factor by which IC~342 is overestimated. We stress that this does not indicate that the BK05 equipartition estimates of IC~342 are necessarily a factor of $\sim$2-3 overestimated, but instead that given reasonable and motivated CR properties and magnetic fields, this overestimation from the BK05 equipartition assumption is easily attainable.

\begin{figure}[t]
\centering
\includegraphics[width=3.35in]{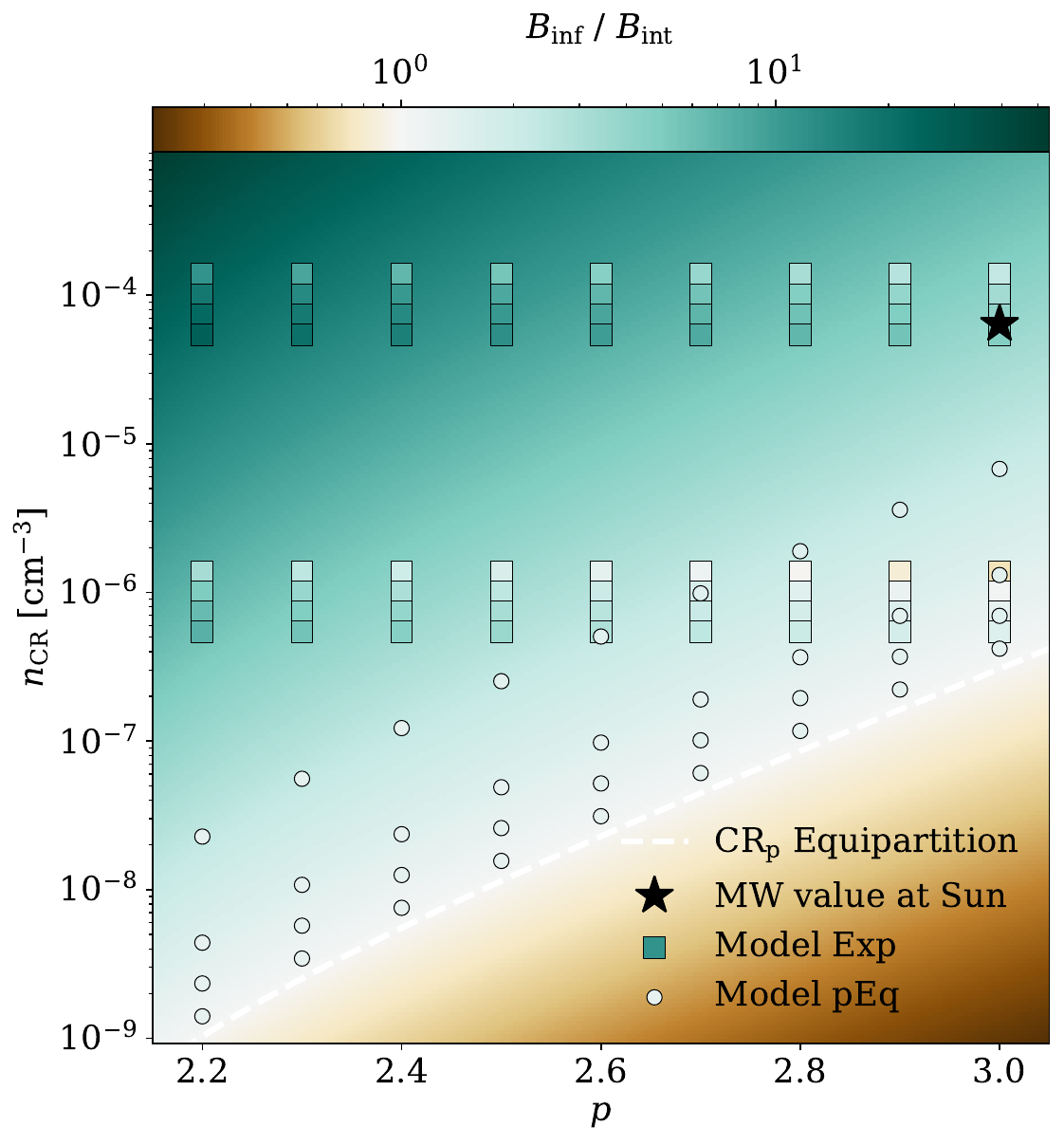}
    \caption{A summary of the over- and under-estimation of the inferred magnetic field strength as a function of power-law index $p$ and $n_{\rm CR}$. Squares (circles) represent average values of the ratio of inferred to intrinsic magnetic fields $B_{\rm inf}$/$B_{\rm int}$ in four large radial bins for the prescribed exponential (imposed equipartition) CR distributions. We color the background according to the theoretical expectation of this misestimation ratio assuming the average magnetic field of the galaxy (see text) in a 1~kpc thick slab. The white dashed line is the theoretical expectation for the equipartition $n_{\rm CR}$ as a function of $p$ for that same mean magnetic field strength. The black star indicates the modeled Milky Way $n_{\rm CR}$ and $p$ as a reference point. There are large regions of ($n_{\rm CR}$, $p$) parameter space that overestimate $B$ by factors of 10 and even 50.}
    \label{fig:square_comparison}
\end{figure}

\begin{figure*}[t!]
\centering
\includegraphics[width=7in]{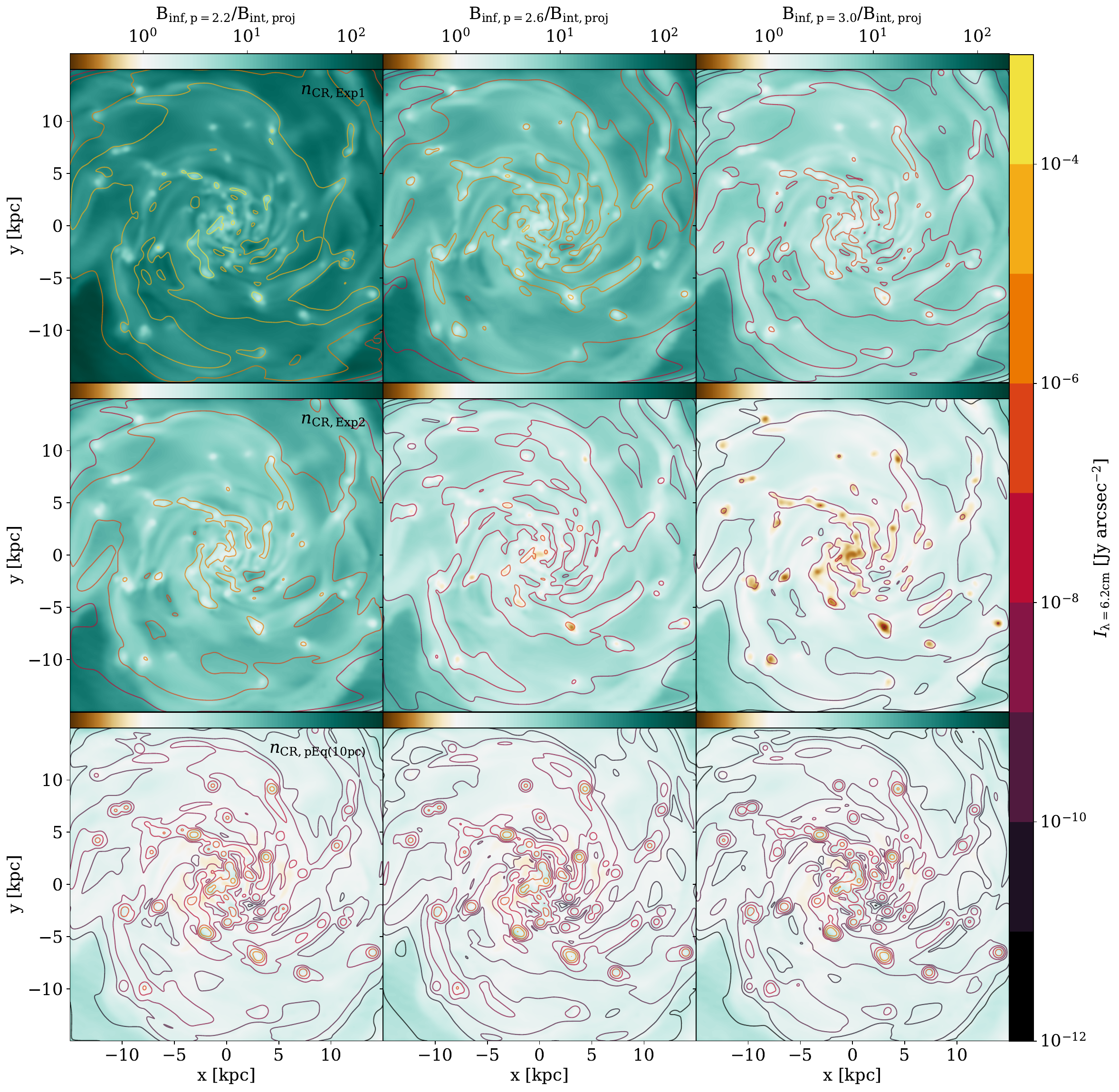}
    \caption{Maps displaying the ratios of inferred (using equipartition) to intrinsic magnetic field strength for three power-law $p$ values: 2.2 (left column), 2.6 (central column), 3.0 (right column) and for our three models of $n_{\rm CR}$. From top to bottom we show the Exp1, Exp2 and pEq (10~pc) models. Overlaid on this ratio map are contours of intensity. For our smooth CR distributions, low intensity inter-arm regions dominate the overestimation of the magnetic field. High intensity regions, which predominantly correspond to dense star-forming clumps, are less overestimated and are even underestimated depending on the CR configuration. As expected, imposed equipartition results in a close recovery of the magnetic field strength, largely independent of the considered regions.}
    \label{fig:ratio_map}
\end{figure*}
We summarize the overestimation (or underestimation) of the inferred galactic magnetic field strengths (from the equipartition assumption) as a function of the power-law index $p$ and the CR electron densities present in the galaxy in \autoref{fig:square_comparison}. The colors on the plot indicate the ratio of the inferred to intrinsic magnetic field $B_{\rm inf}$/$B_{\rm int}$, to quantitatively illustrate how the over- or underestimation varies with $p$ and $n_\text{CR}$. As described for \autoref{fig:intensity} in \autoref{sec:resultsI}, the square and circle points represent radially binned profile values for the two cases of Model Exp and Model pEq(10pc), respectively. In this figure, the profile that is binned in four wide radial bins is the ratio of the inferred to intrinsic profiles. The horizontal and vertical positions of the points are the same as in \autoref{fig:intensity}. The background colors in \autoref{fig:square_comparison} are the theoretical expectation for $B_{\rm inf}$/$B_{\rm int}$ for a given $p$ and $n_\text{CR}$. This ratio is determined using \autoref{eq:polaris_j}, and \autoref{eq:Beck} in a vertical ISM slab 1~kpc thick with a uniform intrinsic $B$. As in \autoref{fig:intensity}, the white line represents the unique $n_\text{CR}$ values defined by CR equipartition in \autoref{eq:nCRdensity_eq_p} for a set of $p$ values and a given intrinsic $B$. The intrinsic $B$ for the theoretical background and equipartition line are set to be the average intrinsic projected magnetic field of the simulated galaxy within 8 kpc (the straight solid lines in \autoref{fig:profiles_B}).

As expected, the theoretical equipartition line exactly traces the theoretical $B_{\rm inf}$/$B_{\rm int}$ of unity, at the center of the white-colored region, as CR electron densities equal to the values necessitated by CR equipartition will exactly recover the provided $B$ field. The same happens for the simulation points with imposed equipartition (circles; see also \autoref{fig:profiles_B}). Both the background prediction and the Model Exp measurements (square points) show large regions of parameter space far from true equipartition, leading to overestimation of the $B$ field by factors of 10 or even 50. This largest overestimation can occur even when the CR densities are set to the values modeled for the Milky Way, particularly at low $p$ and at large radii (lowest square points in the vertical series). For higher $p$ and CR electron densities comparable to those modelled locally for the Milky Way, assuming equipartition can result in an overestimation of $\sim$3 or 4. 

We explore how this misestimation of the magnetic field strength from the application of the equipartition assumption varies spatially across different regions of the galaxy in \autoref{fig:ratio_map}. We show a selection of three representative power-law indices, and our three main CR distribution models. The figure shows maps of $B_{\rm inf}$/$B_{\rm int}$ (where the intrinsic $B$ maps are convolved in the same manner as the intensity maps, see \autoref{sec:synth_obs}). To understand how these ratios relate to the emitted intensity, we overlay on these maps contours of synchrotron intensity.

In the top and central rows, the highest overestimation occurs in lower intensity inter-arm regions. Notably, some of the regions emitting the highest intensities (which also correspond to dense clumps with the highest intrinsic magnetic fields; $B \sim 10^{-4}$ G) either do not overestimate (first row and second row, first two panels) or even underestimate the $B$ field (second row, third column). This effect of the intrinsic $B$ in \autoref{eq:Beck_p} is discussed further in \autoref{sec:B_int_dep}.

For the case of true equipartition, $B_{\rm inf}$/$B_{\rm int}$ is near unity in all regions, as expected. Some inner regions have the magnetic field strength only slightly underestimated whereas in some outer regions it is only slightly overestimated. Minor deviations are attributed to line-of-sight fluctuations in the magnetic field strength.

\autoref{fig:ratio_map} illustrates how, if the distribution of CRs in galaxies is akin to the smooth exponential modeled for the Milky Way (\autoref{eq:nCRdensity}), then star-forming clumps and spiral arms in general are significantly less overestimated (and even underestimated) than low intensity inter-arm regions, where this may reach up to two dex overestimation (depending on $p$). Extreme over- and underestimation could potentially be alleviated by assuming spatially variable $p$, motivated by CR electron energy losses: lower in dense and star-forming regions, and higher in the inter-arms. This is further discussed in \autoref{subsec:spatial_p}.

\begin{figure}[h]
\centering
\includegraphics[width=\columnwidth]{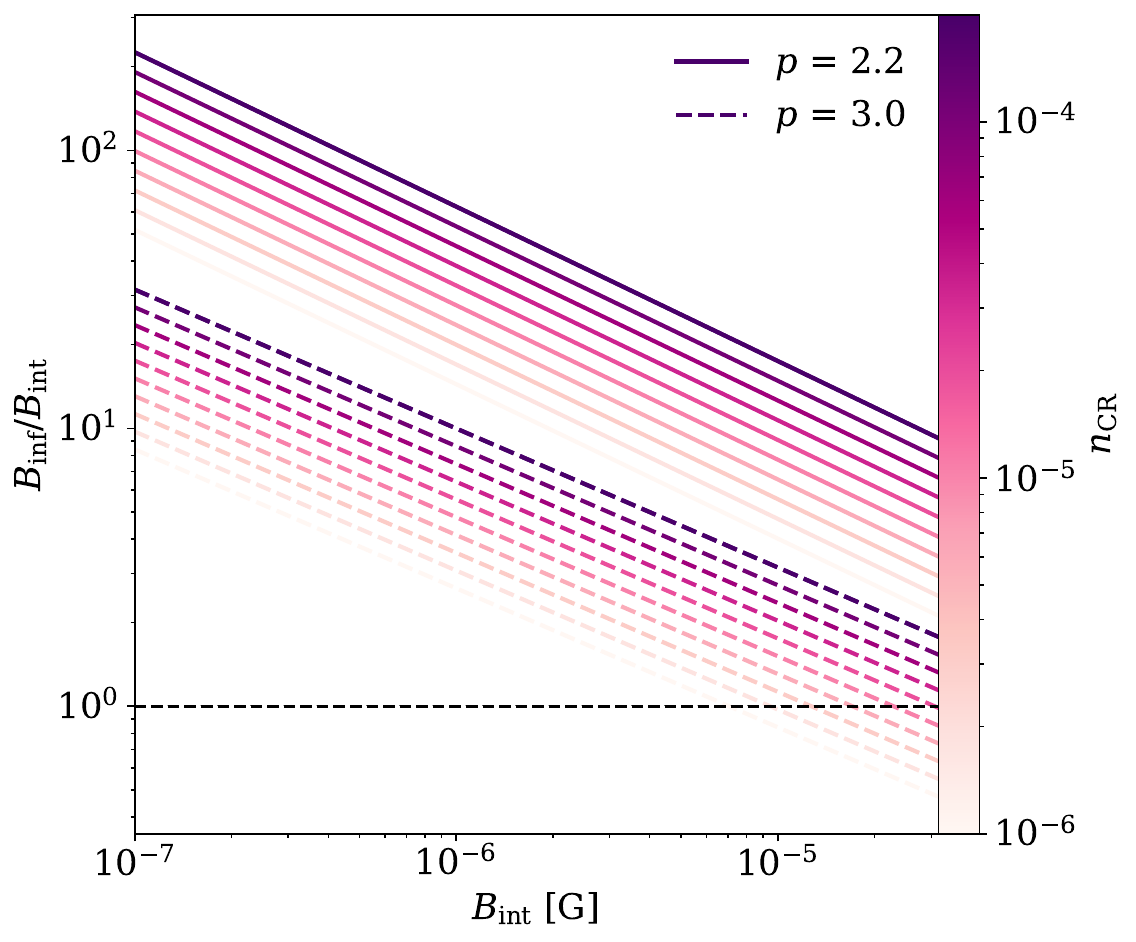}
    \caption{Dependence of the equipartition assumption magnetic field strength misestimation on the strength of the intrinsic magnetic field. We show this for two bracketing power-law indices: $p$ = 2.2 (solid) and $p$ = 3.0 (dashed). A range of $n_{\text{CR}}$ is displayed in light pink to purple for each of the bracketing cases. The intersect with accurate estimates (no overestimation) is marked as a dashed black line at $B_{\rm inf}/B_{\rm int}$ = 1. Higher intrinsic magnetic field strength leads to less overestimation and even underestimation. This plot aids in the interpretation of the above plots and findings for different intrinsic magnetic field strengths and galaxies.}
    \label{fig:B_int_dep_p}
\end{figure}

\subsubsection{Dependence on Intrinsic Magnetic Field}
\label{sec:B_int_dep}
The specific factors of misestimation displayed in \autoref{fig:square_comparison} and \autoref{fig:ratio_map} are necessarily dependent on the intrinsic magnetic fields in the studied system. \autoref{fig:B_int_dep_p} illustrates how the misestimation varies with intrinsic magnetic field $B$, power-law index $p$ (we show only two bracketing cases of $p$ for readability), and CR number density $n_{\rm CR}$. These can be used to guide the intuition of observational inferences of the magnetic field strength. $B_{\rm inf}$/$B_{\rm int}$ (and therefore the factor of over- or underestimation) decreases linearly with increasing $B_\text{int}$. For the wide ranges of $B_{\rm int}$ and $n_\text{CR}$ considered here, overestimation can be as high as 200 and as low as 0.5. The trends are unaffected by assuming different $n_{\rm CR}$ values, and are driven by the dependence on $B$ when assuming an relation between $n_{\rm CR}$ and $B$ through the equipartition assumption in \autoref{eq:polaris_j}. This trend explains the behavior observed for the high intensity clumps in \autoref{fig:ratio_map}, driven by the higher magnetic fields of those regions.

The relation observed for the inferred field with the intrinsic field contextualizes our results and those by \citet{Ponnada2024}. They determine that assuming equipartition, and assuming the emission to arise from the volume-filling phase leads to underestimation of the true magnetic field strengths in the dense gas regions and overestimation in the thick disk and diffuse regions above the disk. This is in broad agreement with our findings in \autoref{sec:results}, and with the physical interpretation we present in \autoref{fig:ratio_map}. \citet{Ponnada2024} also find overall underestimation as opposed to overestimation, resulting from stronger magnetic fields in their simulated galaxies ($\sim15 - 50 \,\mu$G compared to our $\sim2 - 20\,\mu$G). \autoref{fig:B_int_dep_p} shows how depending on the strength of the intrinsic magnetic field, two regimes of estimation are possible: overestimation for lower intrinsic fields and underestimation for stronger ones. This underestimation can also be heightened by the chosen treatment of the lower energy CR spectrum which we discuss in \autoref{subsec:gamma}. While a numerical comparison of MHD schemes and physical models is beyond the scope of this work, we note that stronger magnetic fields ($\sim$$5 - 6\times$ higher within the galaxy; MB10 model) lead to unrealistic galaxy properties \citep{Martin-Alvarez2020} and a lower resemblance to observations \citep{tomography2024} for galaxies simulated with our {\sc ramses} code. Conversely, with our framework of constrained transport MHD simulations paired with flexible post-processing CR electron distributions, we recover realistic flat intensity profiles that resemble those observed in real galaxies (see \autoref{sec:resultsI}), and tend to be in the regime dominated by overestimation of the intrinsic magnetic fields of galaxies when inferring the field strength through the equipartition assumption.

\section{Discussion}
\label{sec:discussion}
In this section, we further examine some important assumptions in our analyses and their implications for our understanding of the intensities produced by imposed CR equipartition and the misestimation induced by the equipartition formulae. 

\subsection{Treatment of the Cosmic Ray Electron Spectrum}
\label{subsec:gamma}

We discuss the treatment of the spectral shape of the CR energy spectrum, and its bounding energies. In this work, we assumed one of the most common assumptions employed to extract information from synchrotron observations: that the energy spectrum of CR electrons is well-described by a power-law (\autoref{eq:nE}). In their integration of this spectrum, BK05 implicitly take its lower bound to be $\gamma_{\rm min} = 2$ (i.e., $E_{\rm min} = E_{\rm e}$). Here, we set the lower bound on our energy integral to be  $\gamma_{\rm min} = 5$ (\autoref{eq:gamma}; i.e., $E_{\rm min} = 4 E_{\rm e}$).

This minimum kinetic energy intervenes in two parts of our analysis, the synchrotron intensity calculated with {\sc polaris}, and the equipartition equation for the CR electron density. The intensity is negligibly affected by this choice of $E_{\rm min}$, as {\sc polaris} corrects for this minimum energy with an additional factor $E_{\rm min}^{1 - p}$ ($\gamma_{\rm min}^{1 - p}$ in their notation; see Appendix A in \citet{Reissl2019}).
Intensities at $\lambda = 6.2\,\cm$ are dominated by the contribution of higher energy electrons due to $\gamma_{\rm min}^2 \ll \lambda/\lambda_{\rm c}$ for $\gamma_{\rm min} = 5$ and below \citep{Reissl2019, Pandya2016}.

\begin{figure*}[t!]
\centering
\includegraphics[width=\textwidth]{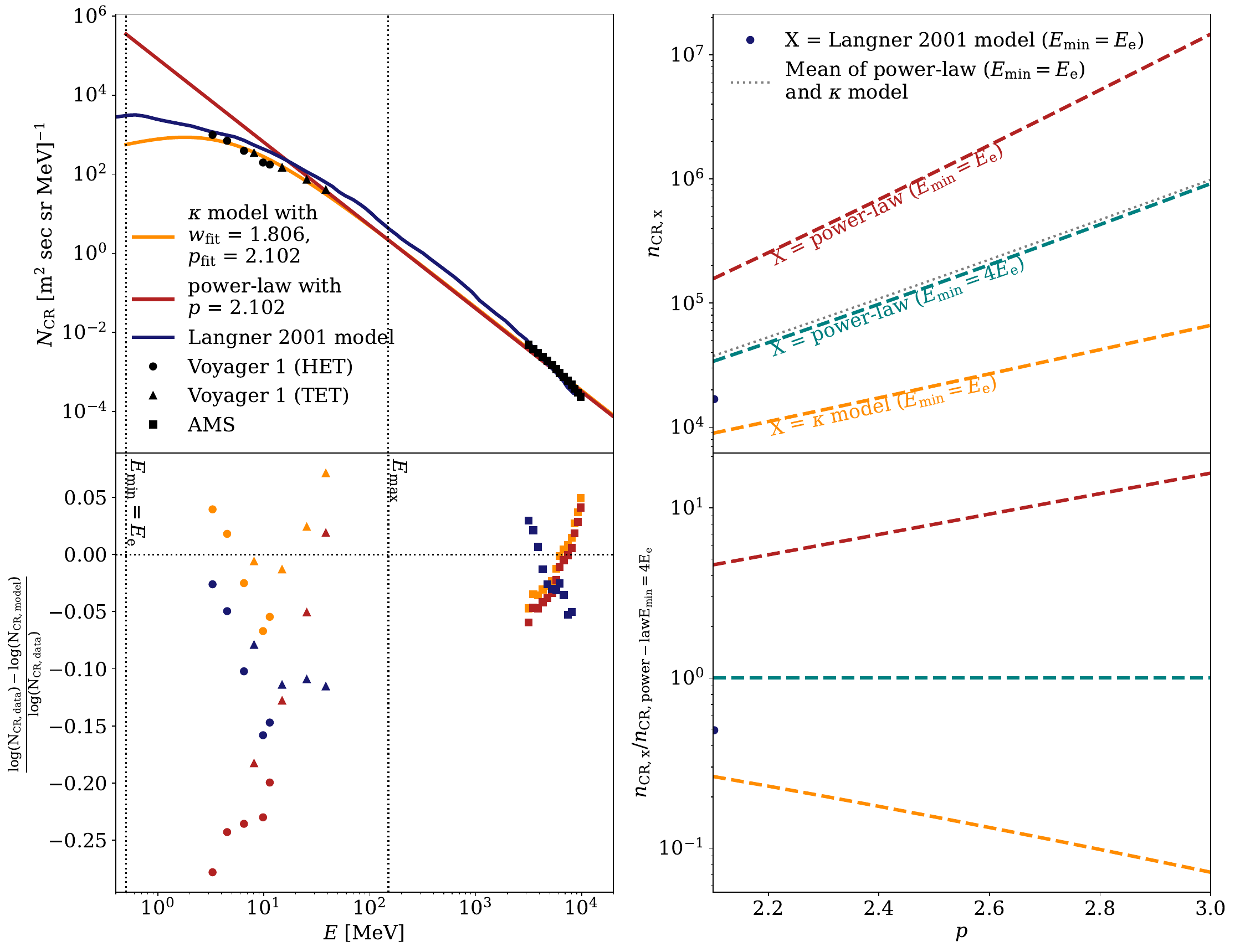}
    \caption{Upper left: comparison between a $\kappa$ and power-law models with the direct CR electron energy spectra measurements \citep[Voyager 1 and AMS data]{Cummings2016}. We fit the $\kappa$ model to the observations, and fix the power-law $p$ to the same high-energy end (see text). We also show a model for the interstellar spectrum from \citet{Langner2001}. Lower left: residuals with respect to the observational data for the $\kappa$ model fit, power-law, and \citet{Langner2001} model. Upper right: integrated $n_\text{CR}$ from the energy spectra between the $E_{\rm min}$ (as indicated in the plot) and $E_{\rm max}$, varying the power-law index $p$ (assuming a fixed $w$ parameter and normalization). Lower right: the ratio of the energy spectrum integrals to the integral of the power-law with a lower bound of $E_{\rm min} = 4 E_{\rm e}$ (teal dashed line). The power-law model for the CR electrons with a lower bound of $E_{\rm min} = 4 E_{\rm e}$ yields an $n_\text{CR}$ nearly identical to the mean of the full power-law and the $\kappa$ models, motivating its use in our analysis as a conservative upper limit for $n_\text{CR}$. The ratios plotted in the lower right panel represent the factors by which alternative choices would modify our CR electron number densities and their resulting intensities.}
    \label{fig:spectrum}
\end{figure*}

The shape of the CR electron energy spectrum is significantly more important for determining the CR electron density resulting from the energy equipartition between the CR protons and the magnetic field (see \autoref{eq:nE} and \autoref{sec:modelpEq}). As the number density of electrons matters in the derivation of the equipartition formulae, it is important to accurately capture the energy contained even within the lower energy end of the CR electron energy spectrum. In \autoref{fig:spectrum}, we show various possible treatments of the energy spectrum $N_{\rm CR} = \frac{dn_{\rm CR}}{dE}$ displayed as a function of the kinetic energy ($E$) to motivate our fiducial choices. The first is the commonly assumed (e.g., BK05 or this work) power-law form of the energy spectrum. A power-law tail towards high energies is highly motivated for such nonthermal distributions, as they are observed in synchrotron-emitting plasmas, solar winds, and particle-in-cell simulations of reconnection and shock acceleration \citep{Pandya2016}. 

However, at lower energies, the abrupt transition from abundant CR electrons to little or none below kinetic energy $E = E_{\rm min} \sim E_{\rm e}$ might be unrealistic. Instead, a rollover to a flatter distribution at lower electron energies is expected due to several mechanisms. If not in the vicinity of the CR acceleration sites, multiple energy loss mechanisms (e.g., Coulomb losses, ionization losses) will lead to shallower slopes at the low energy end of the spectrum \citep{IpAxford1985}. This end might already be flatter after CR electrons are accelerated and propagated through multiple shocks from supernova remnants and a turbulent ISM environment \citep{Bell1978_2, Korpi1999}. Furthermore, these low energy electrons have a lower probability to escape from acceleration sites such as supernovae remnants, leading to a deficit with respect to the power-law scaling \citep{Ohira2012}.

A $\kappa$ distribution function-spectrum \citep{Vasyliunas1968} addresses these problems and captures such a turn-over into a thermal distribution at low energies while maintaining the high $E$ power-law behaviour. This distribution can be expressed \citep{Xiao2006, Pandya2016} as follows: 
\begin{equation}
    N_{\rm CR} \propto \gamma (\gamma^2 - 1)^{1/2} \left( 1 + \frac{\gamma - 1}{(p+1)w}\right)^{-(p + 2)}.
\end{equation}

We also include a model from \citet{Langner2001} that extends to low CR electron energies. This model is based on Galactic polar synchrotron data and was constrained to match high energy electron energy spectrum observations while exceeding the solar modulated Pioneer 10 intensities from the outer heliosphere.

We further include in \autoref{fig:spectrum} observational data for the the CR electron energy spectrum. At the low energy end, we show measurements from Voyager 1 spacecraft between the end of 2012 and 2015, after the spacecraft crossed the heliopause \citep[TET and HET telescopes as circles and triangles, respectively;][]{Cummings2016}. For the high energy end where modulation from the heliopause is minimal, data points correspond to the Alpha Magnetic Spectrometer (AMS) mission on the International Space Station, (\citealt{Aguilar_AMS}; see also \citet{Cummings2016}).

We employ this observational data to fit the $\kappa$ model to the Voyager and AMS data in log-space, resulting to $p = 1.8$ and $w = 2.1$. We fix the displayed power-law to this $p$ index to match the $\kappa$ fit. Both are normalized to the same $N_{\rm CR}$ value at $E_{\rm max}$, well in the power-law regime. A direct fit of the power-law to the observed data results would result in unrealistic $p < 2$ (see BK05 and \citet{Seta&Beck2019} for details on the implications of lower power-law indices), while still failing to capture the high-energy regime.

To illustrate how the $\kappa$ with the low energies turn-over better reproduces the observational data, we show the log-space residuals highlighting the distance between the different models and the observational data in the lower left panel of \autoref{fig:spectrum}. This panel shows how the power-law describes well the high energy behavior but fails to capture the lower energy end, significantly overestimating the number density of CR electrons.

In the derivations of the equipartition CR electron number density, the integral of the CR electron energy spectrum shape at lower energies has significant impact. In the upper right panel of \autoref{fig:spectrum}, we consider $n_{\rm CR,X}$, the integral of each $N_{\rm CR} = \frac{dn_{\rm CR}}{dE}$ curve between $E_{\rm min}$ and an $E_{\rm max} = 300 E_{\rm e}$. We vary the $p$ index to review how the integrals would extrapolate for spectra that have different power-law slopes, but maintain the normalization fixed at $E_{\rm max}$ as well as the $w$ fixed at $w_{\rm fit}$ for simplicity. While the units of $n_{\rm CR}$ are arbitrary here, the ratio of the curves indicates the factor by which these different spectra would change the equipartition CR electron densities, and therefore the intensities resulting from our equipartition models. Importantly, the power-law integrated down to $E_{\rm min} = E_{\rm e}$ leads to significantly higher $n_{\rm CR}$ than the $\kappa$ model integrated to the same $E_{\rm min}$. This is due to the power-law assuming an increasingly large number of electrons all the way down to that minimum energy. We emphasize that the $\kappa$ distribution is a significantly better fit to the lower energy data from direct CR spectrum observations, and that the spectrum is also theoretically expected to deviate from a power-law towards low energies.

By integrating the power-law down to a $E_{\rm min} = 4 E_{\rm e}$, we conduct a conservative treatment of the uncertain energy spectrum integral, as it follows exactly the mean of the power-law and the $\kappa$ model (upper right panel, \autoref{fig:spectrum}). This is emphasized in the lower right panel of \autoref{fig:spectrum} where we take a ratio of those two extreme cases to our power-law model with $E_{\rm min} = 4 E_{\rm e}$. This plot indicates the factors by which our CR electron number densities (and therefore the synchrotron intensities) would be modified for our equipartition models (Model pEq) if a power-law down to $E_{\rm min} = E_{\rm e}$ or a $\kappa$ model fit to the Milky Way observational data were assumed instead. This factor is roughly an order of magnitude higher or lower for those two models respectively. In other words, if the electron spectrum follows the $\kappa$ model fit, the equipartition intensities would be around a factor of 5 - 15 lower in the lowest panel of \autoref{fig:profiles_I}, bringing them to even more tension with the observed intensities of IC~342. If the electron spectrum does indeed continue as a power-law all the way until $E_{\rm min} = E_{\rm e}$ and sharply drops, the equipartition cases will shift upwards by around a factor of 5 - 15, bringing them in better agreement with the observed intensities. Despite this, we note that the profiles will still present an appearance significantly less smooth than that of the observations. Furthermore, the 100~pc equipartition, 1~kpc equipartition, and diffused equipartition to 2 ~kpc would still have intensities considerably lower than those observed. From this, one could assume an uncertainty in our predicted intensities of $\sim$$1$~dex. However, we favour intensities comparable to our teal line model, or even lower as suggested by the $\kappa$ model. Finally, the \citet{Langner2001} model and its integral also favors the use of $E_{\rm min} = 4 E_{\rm e}$ or higher.

Our treatment of the CR electron energy spectrum and our claim that the integral of the spectrum is (conservatively) lower than the integral taken all the way to $E_{\rm min} = E_{\rm e}$ in BK05 is accounted for in our equipartition $B$ field equations through the factor of $E_{\rm min}^{\frac{2(p-1)}{p+5}}$ (see \autoref{eq:Beck_p}).
Along these lines, different assumptions about the energy spectrum would affect the magnetic field recovered using the equipartition assumption (by effectively changing that factor), varying our values for the misestimation of the magnetic field strength by $\sim$$1.5-2$ in our Exp Models. Assuming a realistic $\kappa$ distribution matching the observational measurements would increase overestimation. Alternatively, enforcing a power-law without any spectral flattening down to $E_{\rm min} = E_{\rm e}$ would decrease the overestimation by artificially decreasing the energy budget available for high energy electrons. The above mentioned factors would need to be included in order to take the factors of magnetic field misestimation we have shown to be possible in this work and interpret them through the lens of the BK05 full power-law assumption or the observationally motivated $\kappa$ distribution.

We emphasize that the uncertainty in the low-energy end of the electron spectrum, particularly in other galaxies, and its apparent impact on the magnetic fields inferred through equipartition provide a further reason for caution in its usage. In cases where it is applied, it would also benefit from incorporating alternative spectral distributions (such as the $\kappa$ model) into said assumption.

\subsection{Spatial Variations of the Power-law Index}
\label{subsec:spatial_p}
An assumption that we have made throughout this work is that the power-law index $p$ of the CR electrons and protons (and therefore the synchrotron spectral index $\alpha = (p-1)/2$) remains constant throughout the galaxy. This assumption might be valid on average but it does not generally hold, especially when considering specific sub-regions of galaxies, as discussed in \citet{Beck2007} and \citet{Beck2015}. In particular, the synchrotron spectral index is observed to be lower in star-forming regions, and higher in the inter-arm space. This is likely the result of losses sustained by the CR electrons as they diffuse out from acceleration sites (e.g., \citet{Tabatabaei2007}). Spatial variability in the spectral index has also been observed in the Milky way \citep{SPASS_2018, QIJOTE_2023}. 

By showing our synthetic observations and the comparisons of inferred and intrinsic magnetic field for a range of power-law indices, we avoid fixing to a single $p$ index value. In \autoref{fig:ratio_map}, the reader could interpret the factors of misestimation in the Model Exp cases considering the power-law index they expect in different regions. This may allow the avoidance of extreme over- and underestimation: assuming a lower power-law index in dense star-forming regions and a higher power-law index in the more diffuse inter-arm regions. We caution however that the picture of how and where the power-law steepens and flattens can be more complex when multiple effects are at play. Regions with stronger magnetic fields would also lead to higher synchrotron losses, steepening the local profile with respect to the diffuse regions with weaker magnetic fields. Future work should address in more detail how reasonable models for varying the power-law index spatially affect the misestimation in observed magnetic field strength spatial distributions and radial profiles.

Spatially-varying the power-law index could also affect the amplitude and smoothness of the Model Exp intensity profiles in \autoref{fig:profiles_I}, especially if dramatic deviations from the average profile take place. However, averaging effects may dampen such deviations, particularly under various prescribed CR electron distributions. Note that, due to the negligible variation of the Model pEq intensity profiles with the power-law index (see \autoref{fig:profiles_I}), spatial variations of $p$ are likely to have only a minor effect on the shape of those resulting intensity profiles and therefore would not aid in bringing them to the smoothness of the observations.

While we have assumed a perfect knowledge of the intrinsic power-law index when recovering the magnetic field strength through the equipartition assumption, this is not necessarily the case in observations. The impact of assuming an incorrect $p$ is still to be determined in future work, and motivates new instruments with enhanced resolution that may be capable of recovering resolved synchrotron spectral indices. This encompasses the assumption of a constant power-law index for the entire galaxy when in reality this index actually varies throughout the system.

Finally, while measurements of the CR proton and electron energy spectra by Voyager 1 in \citet{Cummings2016} support the use of the same power-law index for electrons and protons, these data might only be representative of our Local ISM. This is implicitly required by the equipartition formulae presented here, but effects such as e.g., different energy loss mechanisms may invalidate such an assumption as well as the assumptions inherent in $K_0$ (see e.g, \citet{Lacki2013TheLosses} for a discussion of this in the context of starburst galaxies).

The discussed caveats and assumptions indicate that further consideration should be taken into account when measuring magnetic fields from synchrotron observations. In particular, they illustrate how further refinement of the techniques employed for recovering magnetic field information from observations is required. This is particularly urgent in the advent of new facilities such as the Square Kilometer Array (SKA), poised to revolutionize our understanding of astrophysical magnetic fields \citep{SKA_magnetism}.

\section{Conclusions}
\label{sec:conclusions}

In this work we investigate the validity of the common assumption of equipartition in galaxies between the cosmic ray (CR) energy and magnetic energy, in its various forms. We also address the applicability of this assumption in accurately recovering the true magnetic fields intrinsic to galaxies. To do this, we employ constrained transport MHD simulations of a galaxy formed in a high-resolution cosmological zoom-in simulation generated with {\sc ramses}, producing synthetic radio observations of this system using the code {\sc polaris}. We explore how the synchrotron emission from this galaxy varies under multiple configurations of the underlying CR electron distributions. We compare these models with radio observations of the IC~342 galaxy, finding consistency with some of the models while ruling out others, particularly those assuming equipartition. Applying the equipartition assumption to our synthetic observations, we examine the resulting discrepancies between the inferred and intrinsic magnetic fields as a function of specific equipartition assumption, underlying CR electron distribution, CR power-law index $p$, and spatial region in the galaxy. Our main results can be summarized as follows:

\begin{itemize}
    \item Imposing an observationally-motivated smooth $n_\text{CR}$ distribution in our galaxy leads to synthetic synchrotron emission that closely matches the observations. This remarkable resemblance holds for qualitative comparison of intensity maps, and both the smoothness and amplitude of the radial intensity profiles. The agreement of the profile shape is independent of the CR power-law index $p$ which modulates the amplitude alone.
    \item Assuming equipartition between CRs and magnetic energy leads to synchrotron intensities in notable disagreement with observations. These may be ruled out based on the same diagnostics outlined above: intensity maps with emission dominated by dense clumps and the galaxy center, globally low intensities that cannot reach those observed in IC~342, and jagged radial intensity profiles instead of the smoothness characteristic of those from observed galaxies.
    \item We examine various forms of CR-magnetic energy equipartition, and broadly find them unable to match observed intensities. This holds whether we consider equipartition with the global CR energy (dominated by protons), or with the CR electrons alone. A similar result is found for equipartition enforced at different scales. Considering equipartition at small scales ($\sim 10$~pc) fails to match the smoothness of observations. Considering larger scales (100~pc, 1~kpc, or equipartition at small scales with CR diffusion to larger scales) somewhat alleviates the lack of smoothness, but further reduces the overall intensity by de-correlating $n_\text{CR}$ from peaks of high magnetic field strength. 
    \item We test the ability of the equipartition assumption applied to synchrotron observations to recover the true intrinsic magnetic field of the simulated galaxy. For our two exponential CR distributions and at high CR power-law index $p$, the assumption allows the recovery of the approximate order of magnitude of the magnetic field, albeit with a systematic overestimation of the magnetic field strength. This discrepancy is more pronounced at larger radii, as the inferred magnetic field profile presents a different radial shape than the intrinsic one. The degree of overestimation increases with lower $p$ values and increasing $n_\text{CR}$. This unreliable recovery of the inferred field strength is because the same synchrotron emission can be generated by multiple combinations of $B$ and $n_\text{CR}$, with the equipartition assumption fixing them to a specific, but not necessarily correct, pair.
    \item In addition to $p$ and $n_\text{CR}$, the degree of overestimation varies with the strength of the intrinsic magnetic field of the galaxy (see \autoref{fig:B_int_dep_p}). As a result of this dependency, we show how inferred magnetic fields are less overestimated (or even underestimated) in the dense clumps and star forming regions that emit higher intensities. Conversely, we find that magnetic fields inferred for the diffuse ISM and inter-arm regions of our galaxy can be considerably overestimated.
\end{itemize}

There are key theoretical and observational uncertainties broadly associated with CR properties and configurations. The uncertainties on the lower energy end of the CR electron energy spectrum are of particular importance to the questions probed in this work. We employ a conservative treatment of this spectrum, and discuss the details and potential variations to our methods and results in \autoref{subsec:gamma}. We emphasize that the factors of over- and underestimation discussed in this work are with respect to this specific treatment of the equipartition assumption, and are scaled-up by a constant factor of 1.5-2 relative to the misestimation ratios that would be derived for BK05 equipartition magnetic field estimates in the literature (see \autoref{subsec:gamma}). The sensitivity of magnetic field strength estimates to these uncertainties about the cosmic rays motivates further caution with the application of the equipartition assumption.

While additional considerations of the synchrotron emission processes and a detailed and self-consistent treatment of the CR electron spectrum should be further addressed in the future, our findings illustrate how the complex interplay of CR number density, CR energy spectrum, intrinsic magnetic field, and spatial region conspires to impact inferred magnetic field strengths when applying the energy equipartition assumption. With the upcoming Square Kilometer Array, radio astronomy is poised to confront an influx of exquisite and unprecedented new data \citep{SKA_magnetism}. There is a pressing need for new numerical simulations and theoretical infrastructure that are capable of grounding this advent of revolutionary observations, as they bring forth a new era for our understanding of magnetic fields in galaxies and our Universe.

\section{Acknowledgments}
We would like to thank Rainer Beck for helpful discussions and feedback during the writing of this work as well as for providing observational data for IC~342.
The studied simulations were generated at the DiRAC Complexity system, operated by the University of Leicester IT Services, which forms part of the STFC DiRAC HPC Facility (\href{www.dirac.ac.uk}{www.dirac.ac.uk}). This equipment is funded by BIS National E-Infrastructure capital grant ST/K000373/1 and STFC DiRAC Operations grant ST/K0003259/1. The equipment was funded by BEIS capital funding via STFC capital grants ST/K000373/1 and ST/R002363/1 and STFC DiRAC Operations grant ST/R001014/1. DiRAC is part of the National e-Infrastructure.
The computational analysis for this work was performed on the Sherlock cluster using the Oak storage system. We would like to thank Stanford University and the Stanford Research Computing Center for providing these computational resources and support that contributed to these research results.
E.L.-R. and S.M.-A. are supported by the NASA/DLR Stratospheric Observatory for Infrared Astronomy (SOFIA) under the 08\_0012 Program. SOFIA is jointly operated by the Universities Space Research Association,Inc.(USRA), under NASA contract NNA17BF53C, and the Deutsches SOFIA Institut (DSI) under DLR contract 50OK0901 to the University of Stuttgart.
E.L.-R. and S.M.-A. are supported by the NASA Astrophysics Decadal Survey Precursor Science (ADSPS) Program (NNH22ZDA001N-ADSPS) with ID 22-ADSPS22-0009 and agreement number 80NSSC23K1585.
S.M.A. also acknowledges support from the Kavli Institute for Particle Astrophysics and Cosmology (KIPAC) Fellowship. S.E.C. acknowledges support from NSF award AST-2106607, NASA award 80NSSC23K0972, and an Alfred P. Sloan Research Fellowship.

%






\appendix

\begin{figure*}[ht]
    \includegraphics[height = 3.7in]{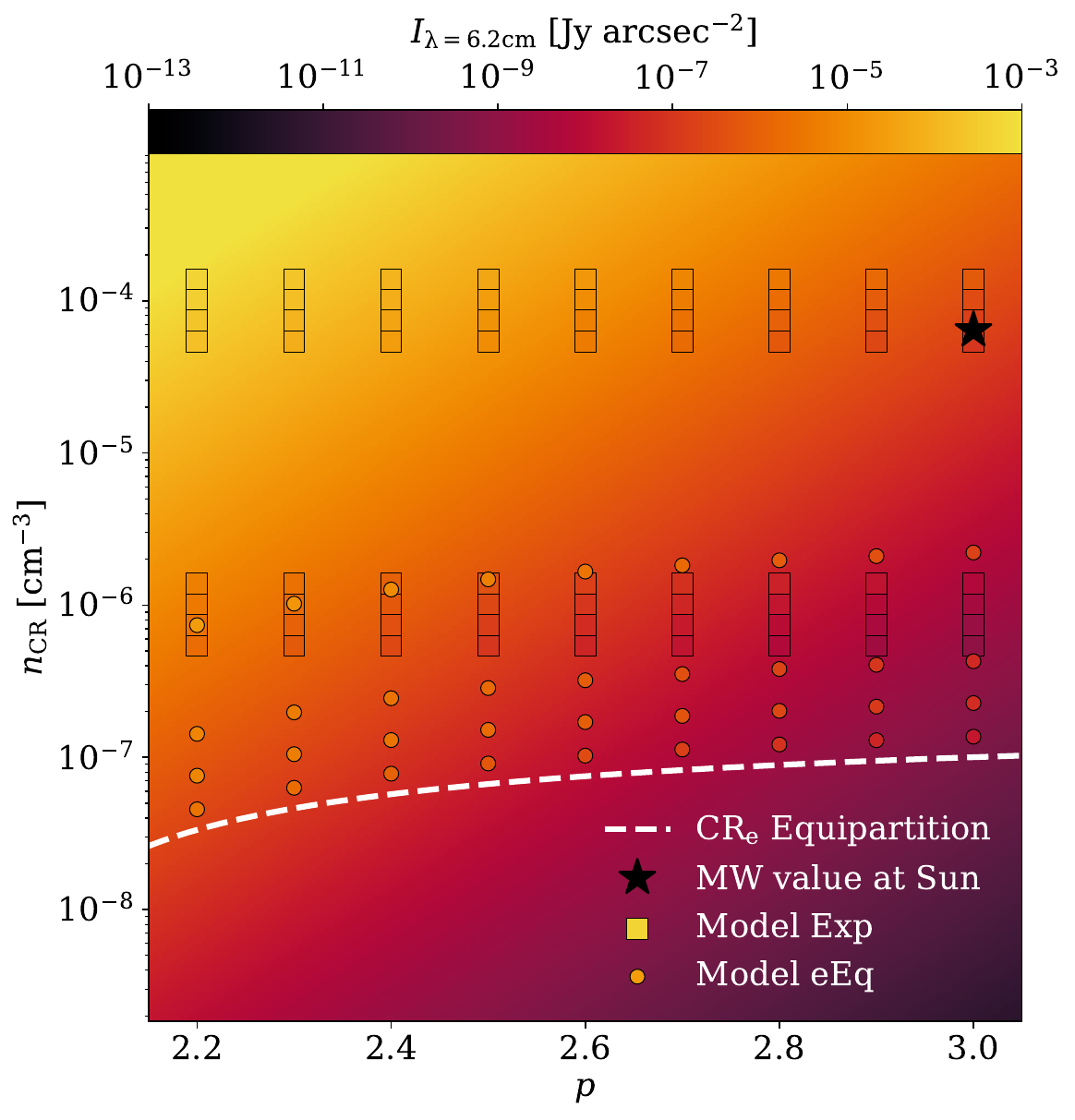}
    \hfill
    \includegraphics[height=3.7in]{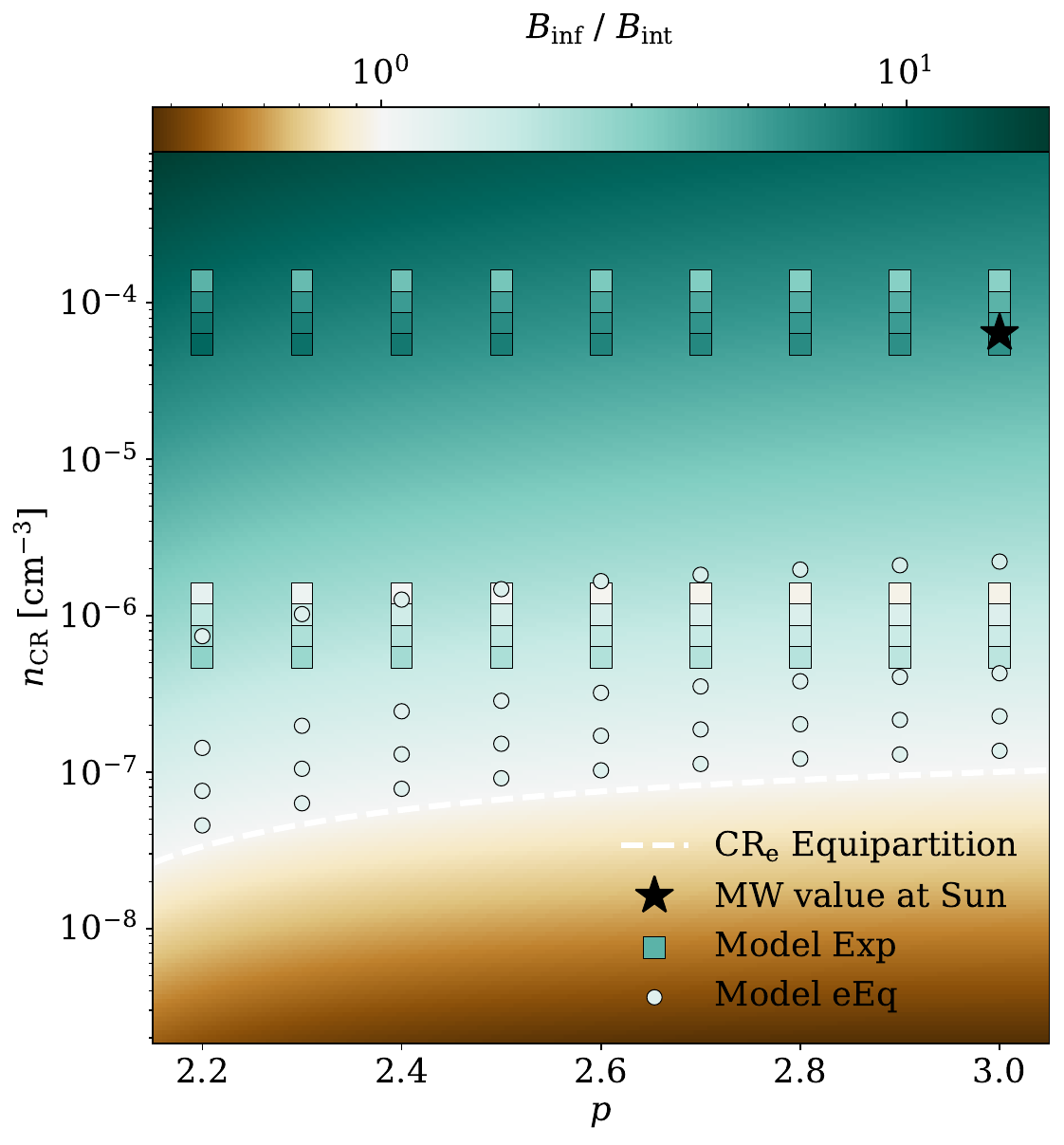}
    \caption{(Left) A summary of the intensity information in \autoref{fig:profiles_I} as a function of power-law index $p$ and $n_{\rm CR}$ with squares (or circles) representing average profile values in four large radial bins for the prescribed (or electron equipartition) CR distributions. The background shows the theoretical expectations of intensities using the mean projected magnetic field strength within 8 kpc and the formula for synchrotron intensity (\autoref{eq:polaris_j}). The white line is the theoretical expectation for the electron equipartition $n_{\rm CR}$ values as a function of $p$ for that same mean magnetic field strength. The black star indicates the modeled Milky Way $n_{\rm CR}$ and $p$ as a reference point. (Right) A summary of the overestimation of the inferred magnetic field strength as a function of power-law index $p$ and $n_{\rm CR}$. The background shows the theoretical expectations of this overestimation ratio using the mean projected magnetic field strength within 8 kpc. The white dashed line and black star are identical to the left figure. The $p$ dependence in electron equipartition is visibly less than with proton equipartition (see \autoref{fig:square_comparison}).}
    \label{fig:square_e}
\end{figure*}

\section{Electron equipartition results}
\label{sec:Results_e}

Having discussed the results of Model pEq (proton equipartition defined in \autoref{eq:nCRdensity_eq_p}) in detail, we briefly present the results of Model eEq (electron equipartition defined in \autoref{eq:nCRdensity_eq_e}) motivated by the equipartition treatment of \citet{Reissl2019} and for completeness. In this model, the CR electrons themselves are directly in energy equipartition with the magnetic field. This is a distinct assumption from that used by BK05 and observational spiral galaxy works in obtaining equipartition magnetic field formulae. As discussed in \autoref{sec:modeleEq}, this electron equipartition is more typically applied to the pair plasmas of relativistic jets and in supernovae sources that are potentially dominated by electrons and positrons and emit synchrotron radiation \citep{Burbidge1956, Chevalier1998, Beck2005RevisedObservations, Duran2013}. Though it might not be particularly motivated for use in the ISM of spiral galaxies, any form of equipartition remains unconstrained observationally, and therefore we test the ramifications of this less common equipartition assumption as well.

\subsection{Synthetically Observed Intensities (with Model eEq)}

The left plot of \autoref{fig:square_e} summarizes the simulated galaxy intensity profiles as a function of power-law index $p$ and CR number density for our Exp and eEq models, as well as the theoretical expectations in the background colors. It is structured identically to \autoref{fig:intensity} and we refer readers to \autoref{sec:resultsI} for a detailed description of the figure layout. 

The theoretical background colors as well as the square points in the left plot of \autoref{fig:square_e} are identical to \autoref{fig:intensity} as the electron equipartition assumption does not play a role in the intensity calculation for a prescribed $n_\text{CR}$. The circular points and the white line, however, reflect the different equipartition assumption and display a significantly flatter relation between $n_{\text{CR}}$ and $p$ as compared to proton equipartition in \autoref{fig:intensity}. Particularly at higher $p$, the region of equipartition (as seen in the circular points and white line) is significantly displaced from the region of reasonable and observed galactic intensities between roughly $10^{-7} - 10^{-4}$ Jy/arcsec$^{-2}$. This distance between observations and electron equipartition is more drastic than with proton equipartition at high $p$ but slightly less drastic at lower $p$ values due to the shallower relation of $n_{\text{CR}}$ with $p$ in electron equipartition. 

\subsection{Intrinsic vs Inferred Magnetic Field (with Model eEq)}
\label{sec:infvint_e}

The right plot of \autoref{fig:square_e} summarizes the $B_{\rm inf}$/$B_{\rm int}$ profiles of the simulated galaxy as a function of power-law index $p$ and CR density distribution, as well as the theoretical expectations in the background colors. It is structured identically to \autoref{fig:square_comparison}. As expected, the circular points that indicate imposed electron equipartition in the $n_{CR}$ have colors implying a $B_{\rm inf}$/$B_{\rm int}$ of near unity (electron equipartition $B$ in \autoref{eq:polaris_B} recovers the true $B$ when the synchrotron emission is generated assuming electron equipartition). The theoretical background as well as the square points show that varying $p$ has a very weak effect on $B_{\rm inf}$/$B_{\rm int}$. This is expected due to the different dependency of the CR electron energy density on the power-law index as compared to the CR proton energy density (\autoref{eq:ep}). They also suggest that smooth CR electron distributions like those modeled for the Milky Way are well out of the regime of electron equipartition and that assuming electron equipartition results in overestimation of the inferred magnetic field strength by around an order of magnitude. Lower CR electron densities (Model Exp2) still lead to overestimation but are significantly closer to the low equipartition $n_\text{CR}$ and therefore give the appearance of recovering the intrinsic field better. We stress that this does not imply the accuracy of the electron equipartition assumption or the low density $n_\text{CR}$ distribution. The true $n_\text{CR}$ is unknown beyond local galactic measurements (see \autoref{sec:discussion}), and we have shown that electron equipartition produces unrealistically low intensities due to its low imposed $n_\text{CR}$. Instead, this simply illustrates how the low density $n_\text{CR}$ of Model Exp is similar to the low densities necessitated by equipartition, allowing electron equipartition to slightly better recover the magnetic field strengths if Model Exp2 was the true $n_\text{CR}$ distribution of galaxies. We emphasize that Model Exp2 is two orders of magnitude lower than the CR electron density deduced for the Milky Way \citep{Sun2008}.  

The spatial distribution of misestimation (not shown) preserves the same qualitative behavior as in \autoref{fig:ratio_map}, but reflects a flatter dependence on $p$. In the case of a smooth exponential distribution of CRs of \autoref{eq:nCRdensity}, the star-forming clumps and spiral arms in general are significantly less overestimated (and even underestimated) than low intensity inter-arm regions. Observational analyses concerned with star-forming regions or spiral arms may apply the electron equipartition magnetic field formula with a single $p$ assumed or measured for the galaxy and retrieve a less overestimated magnetic field strength. This comes at a price of inter-arm regions and the more diffuse ISM featuring measurements overestimated by one or even two orders of magnitude.

\begin{figure}[h]
\centering
\includegraphics[width=\columnwidth]{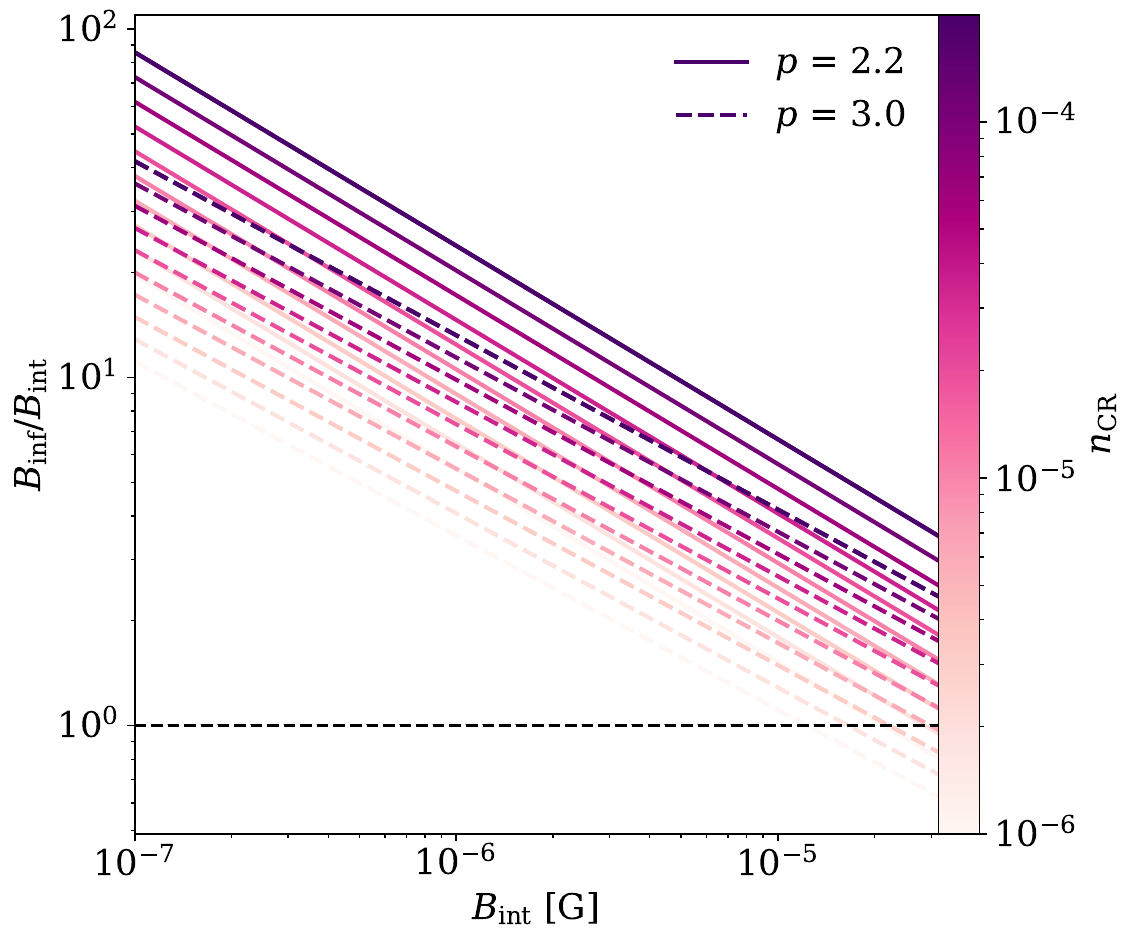}
    \caption{Dependence of the electron equipartition assumption magnetic field strength misestimation on the strength of the intrinsic magnetic field. See \autoref{fig:B_int_dep_p} for details. Higher intrinsic magnetic field strength leads to less overestimation.}
    \label{fig:B_int_dep_e}
\end{figure}

To understand how the electron equipartition overestimation varies with intrinsic $B$, we present \autoref{fig:B_int_dep_e}. Similarly to \autoref{fig:B_int_dep_p}, $B_{\rm inf}$/$B_{\rm int}$ (and therefore the factor of over- or underestimation) decreases linearly as the intrinsic $B$ increases. The bracketing cases of $p$ are significantly closer together with electron equipartition than with proton equipartition. This reflects the weak $p$ dependence seen in the right plot of \autoref{fig:square_e}. As a result, stronger intrinsic $B$ fields would be less overestimated at lower $p$ using electron equipartition than with proton equipartition. This figure provides a framework for observational analysis to estimate the degree of overestimation they might be facing when applying the assumption of energy equipartition between the CR electrons and the magnetic field.

\bibliography{references.bib}{}
\bibliographystyle{aasjournal}



\end{document}